\documentclass[numberedappendix]{emulateapj}
\usepackage{aas_macros}
\usepackage{float}
\usepackage{amsmath,amssymb}
\usepackage{natbib}   
\usepackage{hyperref} 
\usepackage{graphicx} 
\usepackage{color}
\usepackage{verbatim}

\newcommand{\figdir}{Figures}
\newcommand{\proptosim}{\mathrel{\vcenter{
 \offinterlineskip\halign{\hfil$##$\cr
 \propto\cr\noalign{\kern2pt}\sim\cr\noalign{\kern-2pt}}}}}

\newcommand{\unit}[1]{{\rm #1}}

\newcommand{\mean}[1]{\langle #1\rangle}
\newcommand{\method}{L. Wang 2017, in preparation}
\renewcommand{\min}{\mathrm{min}}
\renewcommand{\max}{\mathrm{max}}
\newcommand{\au}{\mathrm{AU}}
\newcommand{\cm}{\unit{cm}}

\newcommand{\g}{\unit{g}}

\newcommand{\K}{\unit{K}}
\newcommand{\km}{\unit{km}}
\newcommand{\kms}{\unit{km\ s^{-1}}}

\newcommand{\erg}{\unit{erg}}
\newcommand{\eV}{\unit{eV}}
\newcommand{\keV}{\unit{keV}}
\newcommand{\s}{\mathrm{s}}
\newcommand{\yr}{\mathrm{yr}}
\newcommand{\hr}{\mathrm{hr}}
\newcommand{\ang}{\ensuremath{\mathrm{\AA}}}
\newcommand{\dyn}{\mathrm{dyn}}

\newcommand{\lya}{\text{Ly}\ensuremath{\alpha~}}
\newcommand{\nad}{\ensuremath{\nabla_\text{ad}}}
\newcommand{\B}{\mathrm{B}}
\newcommand{\kb}{k_\mathrm{B}}

\renewcommand{\d}{\mathrm{d}}
\newcommand{\e}{\mathrm{e}}

\newcommand{\eff}{\mathrm{eff}}
\newcommand{\rr}{\ensuremath{\mathrm{rr}}} 
\newcommand{\evap}{\mathrm{evap}}
\newcommand{\tot}{\mathrm{tot}}
\newcommand{\wind}{\mathrm{wind}}
\newcommand{\amb} {\mathrm{amb}} 
\newcommand{\dust}{\mathrm{dust}}

\newcommand{\crit}{\mathrm{crit}}
\newcommand{\euv}{\mathrm{EUV}}
\newcommand{\he}{\mathrm{HE}}
\newcommand{\xray}{\mathrm{X}}
\newcommand{\ram}{\mathrm{ram}}
\newcommand{\ad}{\mathrm{ad}}   
\newcommand{\eq}{\mathrm{eq}}   
\newcommand{\iso}{\mathrm{iso}} 
\newcommand{\fid}{\mathrm{fid}} 
\newcommand{\rcb}{\mathrm{rcb}} 
\newcommand{\kh}{\mathrm{KH}}   
\newcommand*\chem[1]{\ensuremath{\mathrm{#1}}}

\begin{document}

\title{Evaporation of Low-Mass Planet Atmospheres:
  Multidimensional Hydrodynamics with Consistent
  Thermochemistry}

\author{Lile Wang$^1$, Fei Dai$^{1,2}$}

\footnotetext[1]{Princeton University Observatory,
  Princeton, NJ 08544}

\footnotetext[2]{Department of Physics and Kavli Institute
  for Astrophysics and Space Research, Massachusetts
  Institute of Technology, Cambridge, MA 02139}

\begin{abstract}
  Direct and statistical observational evidences suggest
  that photoevaporation is important in eroding the
  atmosphere of sub-Neptune planets. We construct full
  hydrodynamic simulations, coupled with consistent
  thermochemistry and ray-tracing radiative transfer, to
  understand the physics of atmospheric photoevaporation
  caused by high energy photons from the host star. We
  identify a region on the parameter space where a
  hydrostatic atmosphere cannot be balanced by any plausible
  interplanetary pressure, so that the atmosphere is
  particularly susceptible to loss by Parker wind. This
  region may lead an absence of rich atmosphere
  (substantially H/He) for planets with low mass
  ($M\lesssim 3~M_\oplus$). Improving on previous works, our
  simulations include detailed microphysics and a
  self-consistent thermochemical network. Full numerical
  simulations of photoevaporative outflows shows a typical
  outflow speed $\sim 30~\km~\s^{-1}$ and
  $\dot{M}\sim 4\times 10^{-10}~M_\oplus~\yr^{-1}$ for a
  $5~M_\oplus$ fiducial model rocky-core planet with
  $10^{-2}$ of its mass in the atmosphere. Supersonic
  outflows are not quenched by stellar wind ram pressure (up
  to 5 times the total pressure at transonic points of the
  fiducial model). The outflows modulated by stellar wind
  are collimated towards the night side of the planet, while
  the mass loss rate is only $\sim 25\%$ lower than the
  fiducial model.  By exploring the parameter space, we find
  that EUV photoionization is most important in launching
  photoevaporative wind. Other energetic radiation,
  including X-ray, are of secondary importance. The leading
  cooling mechanism is ro-vibrational molecular cooling and
  adiabatic expansion rather than recombination or \lya
  cooling. The wind speed is considerably higher than the
  escape velocity at the wind base in most cases, hence the
  mass loss rate is proportional to the second power of the
  EUV photosphere size $R_\euv$, instead of the third, as
  suggested by previous works. By calculating the
  evaporation timescale for a grid of planet models, we find
  a significant reduction in the atmosphere for planets with
  $M\lesssim 6~M_\oplus$. We then propose a semi-empirical
  scaling relation for mass loss rate as a function of high
  energy irradiation, planet mass and envelope mass
  fraction, with error mostly $\lesssim 20~\%$. This allows
  us to reproduce the observed bimodal radius distribution
  of sub-Neptune Kepler planets semi-quantitatively.
\end{abstract}


\section{Introduction}
\label{sec:intro}

Kepler has revealed a new class of planets with radii less
than 4 $R_{\oplus}$ and often with orbital distance closer
than those of Mercury \citep[e.g.][]{2011ApJ...736...19B,
  2014ApJS..210...20M}. These planets, colloquially known as
the sub-Neptunes/super-Earths, were shown to be dominant
outcome of planet formation despite that there is no
analogue in our solar system. The ambiguity in the
nomenclature highlights our ignorance about the composition
of these planets. Are they terrestrial planets with a
tenuous atmosphere or are they icy giants with a thick H/He
envelopes? Over the past few years, radial velocity
measurements and transit timing variation analysis have
revealed the masses of $\sim$ 80 of these sub-Neptunes. It
has been suggested by \citet{2015ApJ...801...41R} that 1.6
$R_{\oplus}$ represents a transition radius where planets
smaller than this radius are predominantly rocky, whereas
planets larger than this radius either contain significant
volatiles such as water or are enclosed in a substantial
H/He envelopes. Are planets formed with such a division in
composition? Or were they sculpted by various processes
during the evolution? In this work, we investigate the
influence of photoevaporation in shaping the observed
properties of sub-Neptunes.

When the planet gravitational potential well is not too deep
($M\lesssim 60~M_\oplus$,
e.g. \citealt{2012MNRAS.425.2931O}), incident irradiative
photons are able to deposit sufficient energy into gas
particles in the atmosphere, so that they can escape from
the potential well, resulting in obvious photoevaporation
\citep[see also][] {2003ApJ...598L.121L,
  2004Icar..170..167Y, 2005ApJ...621.1049T,
  2007A&A...461.1185L, 2011A&A...532A...6S}. Direct
evidences of planet photoevaporation have been
reported. These include ``hot Jupiters''
\citep[e.g.][]{2011A&A...532A...6S}, and sub-Neptune planets
\citep[e.g.][]{2015Natur.522..459E}.  In addition, recent
observation showing bimodal distribution of transiting
planet radii \citep{2017AJ....154..109F,
  2017arXiv170607807D}. This bimodality has been predicted
by \citet{Owen2013} and \citet{2014ApJ...792....1L} as a consequence
of photoevaporation, and further discussed by
e.g. \citet{2017arXiv170510810O} (OW17 hereafter) and
\citet{2017arXiv170600251J}.

Models with hydrodynamics and microphysics have been
constructed to understand the physics of evaporating planet
atmosphere. \citet{2009ApJ...693...23M} (hereafter M-CCM09)
constructed semi-analytic models of photoevaporation,
assuming spherical symmetry, for ``hot Jupiters''.  This
scenario is extended to more complicated dynamics by
including X-ray in \citet{2012MNRAS.425.2931O}.  In this
work, we will also construct a semi-analytic model to help
us develop ideas about the basic physical picture of
photoevaporation. Numerical models have also been
constructed by previous works. \citet{2015ApJ...808..173T}
simulated the evaporation of a model hot Jupiter in three
dimensions, featuring the dynamics of photoevaporation
outflow and interactions with orbital
motion. Two-dimensional simulation in
\citet{2016ApJ...820....3C} focused on the interaction with
the ram pressure of stellar
wind. \citet{2017MNRAS.466.2458C} studied on the eventual
fate of the outflow as an evaporating planet orbits around
the central star. Due to the prohibitive computational costs
to evolve a complete thermochemical network, those models
consist only of a minimal set of reactions, i.e.
photoionization/photodissociation of atomic hydrogen and
\lya cooling, or simply uses a one-to-one mapping of gas
temperature to local ionization parameter by radiation.

The main focus of this work is the combination of consistent
thermochemistry with hydrodynamics for modeling the
photoevaporation of planets. We use a mid-scale chemical
network (24 species, including neutral and singly charged
dust grains) that was proved effective for modeling
thermodynamic processes in protoplanetary disks
\citep[][WG17 hereafter] {2017ApJ...847...11W}. We assume
that the typical components in protoplanetary disks, as the
birthplace of planets, should be similar to a primordial
planetary atmosphere. Hydrodynamics in 2.5-dimensions (with
axisymmetry) is coupled with time-dependent ray-tracing
radiative transfer and thermochemistry in every cell across
the simulation domain. Non-equilibrium processes are treated
properly. Those costly calculations can be finished within a
reasonable ``wall-clock time'' by utilizing the power of
graphics processing units (GPUs hereafter). With these
simulations, we expect to achieve better understanding of
the microphysics and hydrodynamics relevant to
photoevaporation, and probably yield predictions and/or
explanations to observables.

This paper is structured as follows. \S\ref{sec:atmosphere}
presents the static atmosphere model without any irradiation
as the initial conditions of our numerical simulations, and
discusses the implications of those hydrostatic
models. \S\ref{sec:semi-ana-model} construct spherical
symmetric semi-analytic models with minimal thermochemistry,
showing the caveats of them which necessitates proper
numerical simulations. In \S\ref{sec:num-method} we describe
the methods of our numerical
simulations. \S\ref{sec:fiducial-profile} presents the setup
and results of the fiducial model.  In
\S\ref{sec:explore-param} we explore and elaborate the
effects of different physical
parameters. \S\ref{sec:discussions} discusses the
implications and applications of our photoevaporation
models.  \S\ref{sec:summary} concludes and summarizes the
paper. Details of mathematical derivations are provided in
the appendices.

\section{Hydrostatics of planet atmosphere}
\label{sec:atmosphere}

The hydrostatic structures of planet atmospheres are
discussed in this section, as the initial condition and
inner boundary conditions of our further numerical
explorations on photoevaporation. In what follows, we will
use the terms ``atmosphere'' and ``envelope''
interchangeably.

We start with a solid core as the inner supporter and the
source of gravity of the atmosphere. The mass-radius
relation approximately obeys $M_c\propto R_c^4$ \citep[see
also][]{2014ApJ...792....1L}. We adopt the terrestrial mean
density at $1~M_\oplus$,
$\rho(M_c = M_\oplus) = 5.5~\g~\cm^{-3}$, unless specially
noted. Outside the solid core lies the atmosphere. The
atmosphere of a planet has two segments: an adiabatic,
convective interior and an (approximately) isothermal,
radiative exterior \citep[e.g.][]{2006ApJ...648..666R,
  2016ApJ...817..107O, 2016ApJ...825...29G}. Due to incident
radiation from the central star at bolometric luminosity
$L_*$, the temperature of the roughly isothermal exterior
for a planet at semi-major axis $a$ satisfies,
\begin{equation}
  \label{eq:T_eq}
  T \simeq T_\eq = 886~\K\
  \left( \dfrac{L_*}{L_\odot} \right)^{1/4}
  \left(\dfrac{a}{0.1~\au} \right)^{-1/2}\ . 
\end{equation}
Transition from adiabatic to isothermal occurs when
eq. \eqref{eq:T_eq} is equated to the temperature in
eq. \eqref{eq:T_adiabatic}, marked by subscript ``rcb''
(short for ``radiative-convective boundary'') attached to
pertinent physical quantities.

\subsection{Convective (adiabatic) interior}
\label{sec:adiabatic-reg}

The adiabatic equation of state (EoS) of gas reads
$p=K \rho^\gamma$, where $p$ is the gas pressure, $\rho$ the
mass density, $\gamma$ the adiabatic index, and $K$ a
constant related to the specific entropy. We neglect the
self gravity of the envelope.
The temperature and density profiles in the adiabatic layer
are given by,
\begin{equation}
  \label{eq:T_adiabatic}
  \begin{split}
    & T = T_0 \left[ 1 + \beta_\ad
      \left(\dfrac{R_c}{r}-1\right) \right]\ ;\ 
    \rho = \rho_0 \left( \dfrac{T}{T_0}
    \right)^{1/(\gamma-1)};
    \\
    & \beta_\ad \equiv \nad
    \left( \dfrac{G M_c\mu}{R_c \kb T_0}\right)\ ;
    \quad
    \nad \equiv \left(\dfrac{\gamma-1}{\gamma}\right)\ .
  \end{split}
\end{equation}
Here $\mu$ is the (dimensional) mean molecular mass, $G$ the
gravitational constant, $M_c$ and $R_c$ the mass and radius
of the solid core beneath the atmosphere respectively, 
$T_0$ and $\rho_0$ temperature and density at the bottom of
adiabatic atmosphere ($r = R_c$) respectively. $\nad$ is the
adiabatic gradient, and $\beta_\ad$ measures gravitational
binding energy against gas energy: the adiabatic inner
envelope is gravitationally unbounded if $\beta_\ad \leq 1$,
which is not discussed in this paper.

As the total mass of isothermal layer is ill-defined (and
practically small compared to the mass in the adabatic
layer; see \S\ref{sec:isothermal-reg}), we characterize the
envelope mass by the mass of the adiabatic segment, by
integrating from $r=R_0$ to $r_\rcb$,
\begin{equation}
  \label{eq:mass-adiabatic}
  \begin{split}
    & M_\ad = 4\pi \int_{R_c}^{r_\rcb}\rho r^2 \d r
    \\
    & = 4 \pi R_c^3 \rho_\rcb
    \left[\nad \left( \dfrac{G M_c \mu}{R_c \kb T_\eq}
      \right) \right]^{\frac{1}{\gamma-1}}
    \left(\dfrac{\beta_\ad}{\beta_\ad-1}
    \right)^{\frac{3\gamma-4}{\gamma-1}}
    \\
    & \quad \times\
    \mathcal{B}
    \left[
      \left(\dfrac{\beta_\ad-1}{\beta_\ad} \right)x;\,
      \dfrac{3\gamma -4}{\gamma-1},
      \dfrac{\gamma}{\gamma-1}
    \right]_{x=1}^{x=r_\rcb/R_c},
  \end{split}
\end{equation}
where
\begin{equation}
  \label{eq:beta-func}
  \mathcal{B}(x;\,a,b) \equiv \int_0^x\d t\
  t^{a-1}(1-t)^{b-1}\ , 
\end{equation}
is the incomplete beta function.

\subsection{Radiative (isothermal) exterior and model
  calibration}
\label{sec:isothermal-reg}

The isothermal layer the density profile is given by,
\begin{equation}
  \label{eq:rho_isothermal}
  \begin{split}
    & \rho = \rho_\rcb \exp\left[ \beta_\iso \left(
        \dfrac{r_\rcb}{r} - 1 \right) \right] \geq
    \rho_\rcb\, \e^{-\beta_\iso}\ ;
    \\
    & \beta_\iso = \dfrac{G M_c \mu}{r_\rcb \kb T_\eq}\ .
  \end{split}
\end{equation}
$\rho$ converges to a finite constant at
$r\rightarrow \infty$; the integration of gas mass diverges
at infinite radius. We setup the initial atmosphere of a
planet by comparing gas pressure at infinite radius in the
isothermal layer to the ambient pressure of e.g. stellar
wind, nominally ($\tilde{p}$ is the reference parameter
characterizing the magnitude of ambient pressure), 
\begin{equation}
  \label{eq:press-wind}
  p_\amb \simeq \tilde{p}\times 10^{-8}~\dyn~\cm^{-2}\
  \left(\dfrac{a}{0.1~\au}\right)^{-2}\ . 
\end{equation}
If
$p_\infty \equiv (\kb T_\eq \rho_\rcb \e^{-\beta_\iso} /
\mu) < p_\amb$, the pressure in the isothermal segment is
matched to the ambient pressure at a finite radius. In this
case, the mass of the isothermal region, estimated from
$r=r_\rcb$ to where $p=p_\amb$, is generally
$\lesssim 10^{-1}$ of $M_\ad$, and characterizing the mass
of atmosphere with $M_\ad$ is well-defined. If
$p_\infty > p_\amb$, the isothermal envelope evaporates at a
very short timescale by hydrodynamic mechanisms that are
irrelevant to photoevaporation \citep[see
e.g.][]{2016ApJ...817..107O}. The latter will be discussed
in \S\ref{sec:mass-loss}, and the former case is the focus
of \S\ref{sec:num-method},

The radiative nature of the isothermal layer helps us to
calibrate the model parameters. Given $M_c$ and $R_c$, a
pair of $(\rho_0, T_0)$ uniquely determines the hydrostatics
of atmosphere, but those quantities are difficult to map
directly onto the physics. In order to express those crucial
parameters more explicitly, here we follow the scheme in
OW17 with some simplifications. At the radiative-convective
boundary, the adiabatic gradient of $\ln T$ is related to
the luminosity of cooling by continuum of the planet $L$,
\begin{equation}
  \label{eq:lumin-condition}
  \left. \dfrac{\d \ln T}{\d r} \right|_{r_\rcb}
  = -\dfrac{G M_c \mu}{\kb T_\eq r_\rcb^2}
  = -\left(\dfrac{L}{4\pi r_\rcb^2} \right)
  \left( \dfrac{3\kappa \rho_\rcb}{16 \sigma T_\eq^4}
  \right)\ ,
\end{equation}
where $\sigma$ here is the Stefan-Boltzmann constant, and
$\kappa$ is the Rosseland mean opacity. We adopt the fitting
formula of $\kappa$, as a function of $\rho$ and $T$
(obtained by \citealt{2010ApJ...712..974R}, based on
\citealt{2008ApJS..174..504F}),
\begin{equation}
  \label{eq:rosseland-opacity}
  \begin{split}
    & \kappa \simeq \kappa_0 \times \tilde{\kappa}
    \left(\dfrac{p}{\dyn~\cm^{-2}}\right)^\alpha
     \left(\dfrac{T}{\K}\right)^\beta\ ;
    \\
    & \kappa_0 \equiv 10^{-7}~\cm^2~\g^{-1}\ , \quad
    \alpha \equiv 0.45\ ,\quad \beta \equiv 0.68\ .
  \end{split}
\end{equation}
Our ignorance about $L$, as well as the uncertainty in the
dimensionless opacity parameter $\tilde{\kappa}$, are
absorbed into a parameter with the dimension of time,
\begin{equation}
  \label{eq:tau-kh}
  \tau_\kh \equiv \dfrac{G M_c M_\ad}{R_c L \tilde{\kappa}}\ .
\end{equation}
This $\tau_\kh$ is indeed the Kelvin-Helmholtz timescale of
planet atmosphere with $\tilde{\kappa}$ absorbed. From
eq. \eqref{eq:lumin-condition} we have,
\begin{equation}
  \label{eq:m_ad-tau}
  \begin{split}
    & M_\ad  = \dfrac{64\pi \mu \sigma T_\eq^3R_c \tau_\kh}
    {3 \kb (p_\rcb / \dyn\ \cm^{-2})^\alpha
       \kappa_0(T_\eq/\K)^\beta\rho_\rcb }\ ;
    \\
    & \left(\dfrac{\rho_\rcb}{10^{-3}~\g~\cm^{-3}}
    \right)^{\alpha+1}
    \\
    & = \left[10^{-1-3\beta}
      (3.51\times 10^7)^{1-\alpha}\right]\times
    \left(\dfrac{M_\ad}{0.104~M_\oplus}\right)^{-1}
    \\
    &  \times
    \left(\dfrac{R_c}{R_\oplus}\right)
    \left(\dfrac{\tau_\kh}{10^8~\yr}\right)
    \left(\dfrac{\mu}{2.35~m_p}\right)^{\alpha + 1}
    \left(\dfrac{T_\eq}{10^3~\K}\right)^{3-\alpha-\beta}.
  \end{split}
\end{equation}
For fixed $M_c$ and $R_c$, we first take $M_\ad$ and
$\tau_\kh$ as input parameters, then use
eq. \eqref{eq:m_ad-tau} to find out $\rho_\rcb$, before
inserting those quantities into
eq. \eqref{eq:mass-adiabatic} to solve for $\beta_\ad$
numerically, utilizing eq. \eqref{eq:T_adiabatic}. Those
steps allows us to determine atmospheric profiles of the
planet as a function of envelope mass fraction and the
Kelvin-Helmholtz timescale.

\subsection{Mass loss with unbalanced ambient pressure}
\label{sec:mass-loss}

\begin{figure}
  \centering
  \includegraphics [width=3.3in, keepaspectratio]
  {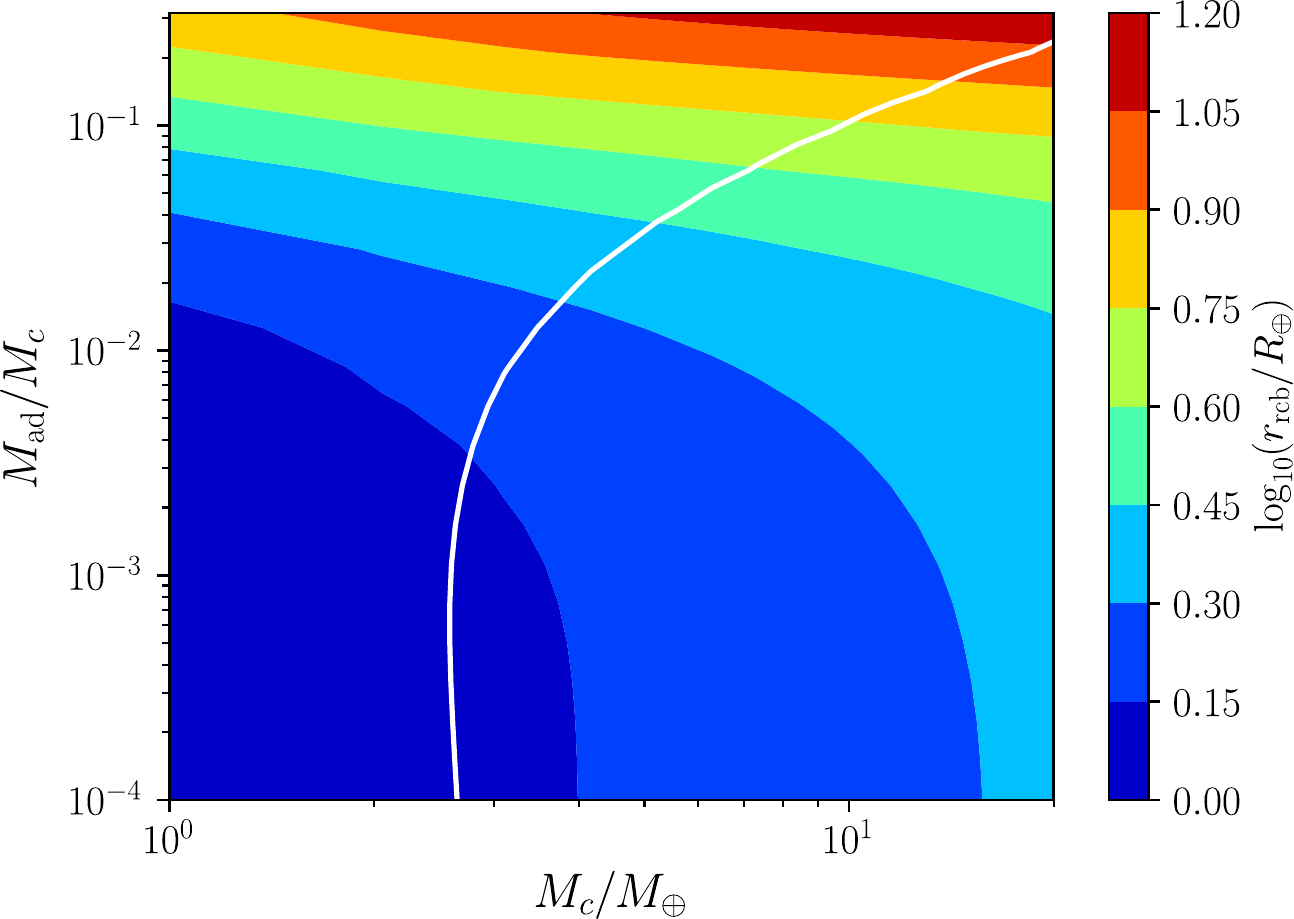}
  \caption{Radius of the radiative-convective boundary
    $r_\rcb$, which roughly resembles the {\it observed}
    radius of planets, presented as a function of solid core
    mass $M_c$ and envelope mass fraction characterized by
    $M_\ad/M_c$ ($M_\ad$ is the mass of the adiabatic
    atmosphere). The white curve overlaid indicates the
    limit at which the ambient pressure can marginally
    confine a hydrostatic atmosphere.  The region on its
    upper left consists of models that would rapidly
    disperse their envelopes due to Parker wind.  Note the
    critical core mass that occurs at $M_c\sim 3~M_\oplus$,
    which depends logarithmically on the luminosity of the
    host star, the core composition, and the ambient
    pressure.}
  \label{fig:r_rcb_example}
\end{figure}

We present examples of hydrostatic models in Figure
\ref{fig:r_rcb_example}. Model parameters are:
$\tau_\kh = 10^8~\yr$, $T_\eq = 886~\K$, $\mu = 2.35~m_p$
($m_p$ is the mass of proton), and $\tilde{p} = 1$. Note the
overlaid white curve indicating the critical conditions of
pressure balancing, above which the models have
$p_\infty > p_\amb$ and thus the atmosphere is expected to
lose mass rapidly. The approximate mass loss rate by Parker
wind mechanism, assuming sufficient energy supply to
maintain an isothermal outflow, is estimated by (subscript
``s'' denotes the sonic surface; for estimating $\rho_\s$
see e.g. \citealt{1958ApJ...128..664P}),
\begin{equation}
  \label{eq:massloss-parker}
  \begin{split}
    &\dot{M}_\mathrm{Parker} = 4 \pi R_\s^2 \rho_\s c_\s\ ;
    \quad 
    R_\s = \dfrac{G M_c}{2 c_s^2} \ ;
    \\
    & \rho_\s \sim
    \rho_\rcb \exp\left[ -\dfrac{R_s}{r_\rcb}
      \left(1 - \dfrac{r_\rcb}{R_\s}\right)^2 \right]\ .
  \end{split}
\end{equation}
On the other hand, the rate of energy injection by the
bolometric luminosity $L_*$ of the host star (not to be
confused with photoevaporation by high energy photons,
$L_\he$) is also limiting hydrodynamic mass loss rate. The
energy-limited mass loss rate $\dot{M}_{\mathrm{ene}}$
is approximately (note that $\pi r_\rcb^2$ is
roughly the area of intercepting stellar radiation in
optical and infrared, and $a$ is the semi-major axis of the
planet orbit),
\begin{equation}
  \label{eq:massloss-ene}
  \begin{split}
    \dot{M}_{\mathrm{ene}} & \sim
    \left(\dfrac{L_*}{4\pi a^2}\right)
    \pi r_\rcb^2 \left( \dfrac{c_s^2}{2} \right)^{-1}\ ;
    \\
    \dot{M} & \sim \min \{\dot{M}_{\mathrm{Parker}},
    \dot{M}_{\mathrm{ene}}\}\ .
  \end{split}
\end{equation}
The evaporation timescale of the adiabatic atmosphere is
approximately
$t_\evap \equiv M_\ad / \dot{M} \gtrsim 10^2~\yr$ for
planets with $M_\ad/M_c\gtrsim 10^{-2}$. Envelope of a
planet evaporates until it reaches the curve if $M_c$ is
greater than the critical value $M_\crit\sim 2.5~M_\oplus$,
or otherwise totally loses its adiabatic segment of
envelope. For rocky-core planets, we varied $\tilde{p}$ by
$\pm 4$ orders of magnitude to confirm that the value of
$M_\crit$ varies within the range of
$2\lesssim (M_\crit/M_\oplus) \lesssim 3$, which depends on
the logarithm (thus very insensitively) of $\tilde{p}$ and
$T_\eq$, as one can infer from eq. \ref{eq:press-wind}.  We
refer the reader to \S\ref{sec:evolution} for examples of
evaporation timescale in this region.

This critical mass quantitatively agrees with the observed
absence of massive H/He planetary atmosphere for less
massive planets with $M\lesssim 3~M_\oplus$ \citep[e.g.][]
{2015ApJ...801...41R, 2016ApJ...819..127Z,
  2016AJ....152..204L}. We hence suggest that the inability
of low mass planets to have its atmosphere pressure balanced
by the ambient would be possible to result in those
observation constraints. Similar mechanisms are also
suggested in \citet{2016ApJ...817..107O} to shape a similar
limit.

This would suggest that planets with
$M_c\lesssim 3~M_\oplus$ would lose its primordial H/He
atmosphere quickly after the disk disperses (after
$10^{6-7}~\yr$). In other words, planets in this mass range
will not have a substantial H/He envelope ($< 10^{-4}$ by
mass) unless subsequent outgassing is
significant. \citet{2016ApJ...817..107O} reached a similar
conclusion but did not offer an quantitative prediction on
the threshold. We will show in \S\ref{sec:discussions}, that
observational test of this prediction is more
complicated. This is because planets with core masses
between $\sim 3~M_\oplus$ to $\sim 6~M_\oplus$ (the upper
boundary is more sensitive to the specific choices of
parameters) may be stable to Parker Wind outflow but still
susceptible to photoevaporation. On a $\sim 10^8~\yr$
timescale, these planets will also lose their primordial
H/He envelope photoevaporatively. Observationally, the
$6~M_\oplus$ (or $\sim 1.6~R_\oplus$) threshold was pointed
out by \citet{2015ApJ...800..135D} and
\citet{2015ApJ...801...41R}.

\section{First Impression on Photoevaporation:
  Semi-analytic Models}
\label{sec:semi-ana-model}

In this section we describe an analytic model with minimum
(but adequate) hydrodynamics, radiative transfer and
thermochemistry, to help us understand the process of
photoevaporation. The procedures are similar to (but still
subtly different from) M-CCM09, which are stated below and
in Appendix \ref{sec:app-semi-ana}.

\subsection{Guiding physics and example solution}
\label{sec:ana-phys}

We consider a spherically symmetric model with external
gravity set by the co-centered solid core. The hypothetical
configuration of stellar radiation is characterized by
radiation flux $\vec{F}=-\hat{r}F_0$ in absence of
absorption. This configuration of radiation field represents
the radial column at the substellar point, and is also
expected to characterize other radial columns
semi-quantitatively. For simplicity we assume that radiation
is monochromatic, and only take the following reactions into
account,
\begin{equation}
  \label{eq:ana-reac}
  \begin{split}
    \chem{H} + h \nu & \rightarrow \chem{H}^+ + e^- \ ;
    \\
    \chem{H}^+ + e^- & \rightarrow \chem{H}\ .
  \end{split}
\end{equation}
At $h\nu = 25~\eV$, the photoionization cross section is
$\sigma \simeq 1.2\times 10^{-18}~\cm^2$
\citep{Verner+etal1996}. We adopt the \verb|UMIST| version
of type-B recombination rate \citep{UMIST2013},
$\alpha_\B \equiv \alpha_0 (T / T_0)^{\kappa}$, where $T_0$
is some fiducial temperature, $\kappa \equiv - 0.75$, and
\begin{equation}
  \alpha_0 = 3.5 \times 10^{- 12} \  \mathrm{cm}^3
  \  \text{s}^{- 1} \  \left( \dfrac{T_0}{300 \ 
  \K} \right)^{\kappa} \  .
\end{equation}
Basic heating and cooling processes corresponding to the two
reactions in eq. \eqref{eq:ana-reac} are also introduced,
including photoelectric heating with energy per reaction
$\langle E_{\mathrm{pe}} \rangle = h \nu - I_e$
($I_e = 13.6 \ \mathrm{eV}$ is the ionization threshold of
atomic hydrogen) for each hydrogen atom ionized, and
recombination cooling energy loss
$\mean{E_\rr} = 3 \kb T / 2$ for each hydrogen atom reformed
($\kb$ is the Boltzmann constant). \citet{DraineBook}
suggests that $\mean{E_\rr}$ is smaller than $3 \kb T / 2$,
since the electrons with lower kinetic energy are easier to
be captured by ions. However, we still assume
$\mean{E_\rr} = 3 \kb T / 2$ for simplicity.

\begin{figure}
  \centering
  \includegraphics [width=3.3in, keepaspectratio]
  {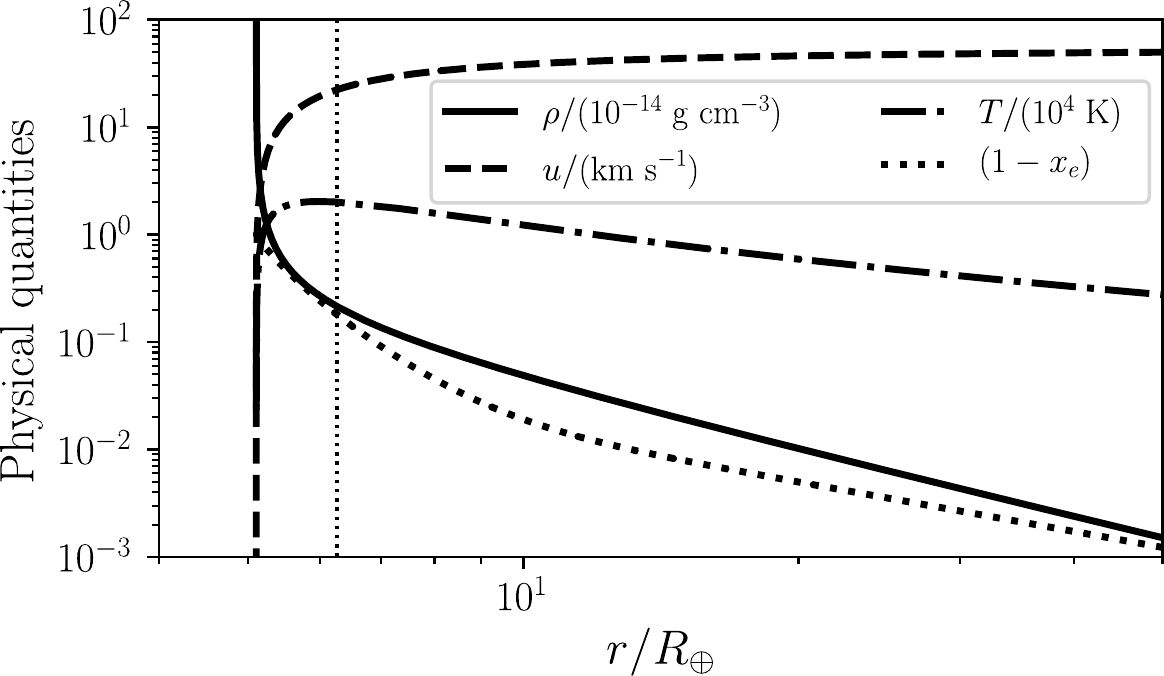}
  \caption{The radial profile of various quantities in the
    spherically symmetric analytical solution of a
    $5~M_\oplus$, $10^{-2}$ envelope mass fraction at
    $0.1~\au$, presented in \S\ref{sec:ana-phys}. Mass
    density $\rho$, radial velocity $u$, temperature $T$,
    and neutral fraction $(1-x_e)$ profiles are presented in
    different line styles, which are indicated in the
    legend. The vertical thin dotted line indicates the
    location where the flow is transonic. }
  \label{fig:ana_example}
\end{figure}

In Figure \ref{fig:ana_example}, we present an example of
semi-analytic solutions. Following the construction
procedures in Appendix \ref{sec:app-semi-ana}, this solution
is matched (a) inwards to a $T_\eq = 886~\K$ isothermal
layer outside an $M_\ad=10^{-2}~M_c$ adiabatic envelope at
$r_\min = 5.11~R_\oplus$ (where
$\rho = 0.98\times 10^{-13}~\g~\cm^{-3}$), surrounding an
$M_c = 5~M_\oplus$, $R_c = 1.5~R_\oplus$ core (assuming
envelope cooling timescale $10^8~\yr$), and (b) outwards to
an $F_0 = 10^{15}~\cm^{-2}~\s^{-1}$ outer boundary (this is
the radiation flux at $a=0.1~\au$ with EUV luminosity
$10^{-3.5}~L_\odot$ at $h\nu=25~\eV$). Considering the
effective solid angle of intercepting EUV irradiation being
$\pi$, the mass loss rate is estimated by
$\dot{M}\sim (\pi r^2 \rho u)_{r_\max}\simeq 1.3\times
10^{-9}~M_\oplus~\yr^{-1}$, thus
$t_\evap\sim 40~\mathrm{Myr}$.

\subsection{Steady versus static: Caveats and necessity of a
  consistent model}
\label{sec:static-caveat}

Straightforward as it may seem, the semi-analytic solution
presented in the previous section still has caveats.  To be
mathematically rigorous, a steady state outflow solution has
no hope to be {\it fully} matched to a static interior: the
radial mass flux, being a finite constant in the steady
outflow, must vanish in the static region. Typically this
mismatch is interpreted by assuming that the static region
quickly relaxes to new configurations as the gas at the wind
base (definition see Appendix \ref{sec:app-semi-ana}) is
removed by the outflow. The new configurations are almost
identical to the original one as long as the evolution time
$t\ll t_\evap$, so that the location and hydrodynamic
conditions of the wind base can be treated as invariant.

However, the equilibrium conditions beneath the wind base
are impossible, with only hydrogen ionization and
recombination processes included. Since
$\mean{E_\mathrm{pe}}\sim (\kb\times 10^5~\K)$ is much
greater than $\mean{E_\rr}\lesssim (\kb\times 10^3~\K)$, in
case of $S_I=0$ (ionization equilibrium), the local energy
balance $S_E$ is always appreciably positive (see also
eqs. \ref{eq:ene-src}, \ref{eq:ionization}). \lya cooling,
suggested by M-CCM09, is indeed negligible at the
temperature and density beneath the wind base for our
low-mass planet models. As a result, the timescale at which
a fluid element in the static region doubles its temperature
is estimated by
$t_\mathrm{heat,static} \sim 10~\s\times
[F/(10^{15}~\cm^{-2}~\s^{-1})]^{-1}$, which is merely one
day even if the EUV flux $F$ is suppressed by 4 orders of
magnitude compared to the unattenuated flux $F_0$.  As soon
as a fluid element is heated, it expands and allows more EUV
radiation to come in, which in turn speeds up the heating
process before this fluid element finally migrates into the
wind. As a result, the eroded static layer never recovers
its original configuration; instead, the location of wind
base moves inwards in a relatively short period of time
compared to $t_\evap$. The wind base shrinks until it
reaches a very high density so that the speed of erosion is
comparable to $t_\evap$. In fact, using the code described
in \S\ref{sec:num-method} in spherically symmetric
configuration and identical thermochemistry as in
\S\eqref{sec:ana-phys}, we figure out that the wind base
moves from $\rho \sim 10^{-13}~\g~\cm^{-3}$ to
$\rho\sim 10^{-6}~\g~\cm^{-3}$ within $t \lesssim 0.1~\yr$,
while the mass loss rate roughly halves as the effective
area of intercepting EUV radiation shrinks (the mass of
isothermal layer in the range of
$10^{-13}~\g~\cm^{-3} < \rho < 10^{-6}~\g~\cm^{-3} $ is only
$\sim 10^{-10}~M_\oplus$). At each instant the wind region
can still be perfectly fit by a semi-analytic solution
(mathematically thanks to the extra degree of freedom, see
the discussions in Appendix \ref{sec:app-semi-ana}; note
that the timescales for a wind configuration to relax is
$r_\max/ v_\wind\sim 10^4~\s$, which is still tiny compared
to the erosion timescale).

In summary, the assumption of quasi-invariant static region
matching a steady state wind with hydrogen only leads to
contradictions.  Therefore we need numerical models where
detailed microphysics are coupled consistently with full
hydrodynamics, preferably in multiple dimensions.

\section{Numerical simulation methods}
\label{sec:num-method}

We present the numerical methods for modeling
photoevaporation of planet atmosphere in this section. Those
schemes are close to WG17 as underlying physical processes
are similar (see \S\ref{sec:intro}).

\subsection{Fluid mechanics}
\label{sec:method-hydro}
 
Full hydrodynamics is included using the general-purpose
grid-based astrophysical simulation code \verb|Athena++|
(\citealt{2016ApJS..225...22W}; J. Stone et al., in
preparation). Despite its capability of solving MHD
problems, we neglect magnetic fields for now, using the HLLC
Riemann solver, van Leer reconstruction with revised slope
limiter for improved order of accuracy
\citep[see][]{2014JCoPh.270..784M}, and Consistent
Multi-fluid Advection (CMA) for strict conservation of
chemicals inside advecting fluids \citep[e.g.][]
{2010MNRAS.404....2G}.

\subsection{Radiative transfer}
\label{sec:rad-trans}

High energy photons from the host star is the key to
photoevaporation. We use four representative energy to
portray the influence of those photons: $h\nu = 7~\eV$ for
FUV photons (``soft FUV'') that do not interact with
hydrogen or helium, $h\nu = 12~\eV$ for the Lyman-Werner
(LW) photons, $h\nu = 25~\eV$ for the EUV photons, and
$h\nu = 1~\keV$ for the X-ray. For high energy photons in
those energy bins, absorption processes overwhelm scattering
\citep{Verner+Yakovlev1995, Verner+etal1996, DraineBook},
with two potential exceptions: hard X-ray and
\lya. Scattered hard X-ray photons could affect ionization
and thus magnetic coupling in regions beyond the reach of
photons in other energy bands \citep[e.g.][]
{Igea+Glassgold1999, Bai+Goodman2009}. We ignore their
effects for two reasons: (a) that magnetic fields are not
included in this paper, and (b) that those scattered X-ray
photons only have marginal thermodynamic impact. \lya do not
deposit appreciable amount of energy into the system,
neither do they destroy \chem{H_2} or \chem{CO}. They do
dissociate \chem{H_2O} and \chem{OH}, which can be important
coolants. Nevertheless, with our numerical experiments, the
soft FUV photons have the same effect but will likely
penetrate deeper. We will revisit the \lya scatter problem
after obtaining the distribution of neutral hydrogen by
simulations.\footnote{\lya cooling is still included in our
  simulations (\S\ref{sec:thermochemistry}; see also WG17).}

Photons of those representative energy bins are traced by
non-radial rays in curve-linear coordinates for the
radiative transfer problem (\method). For photochemistry, we
must calculate the local effective flux for each cell,
\begin{equation}
  \label{eq:eff-flux}
  F_\eff(\nu) = \sum_{\{i\mathrm{\ in\ cell}\}} F_i(\nu)
  \left\{ \dfrac{1 - \exp[-\delta l_i / \lambda(\nu)]}
    {\delta l_i / \lambda(\nu)} \right\}\ ,
\end{equation}
where $F_i$ is the incoming flux on the cell boundary
carried by the $i$th ray, $\delta l_i$ is the chord length
of the ray crossing the cell, and $\lambda(\nu)$ is the mean
free path of photons at frequency $\nu$ with all absorption
mechanisms taken into account, updated along with the
evolution of chemical reaction network. The flux of a ray in
each energy bin is also adjusted as it propagates through
each cell according to the photochemical reactions and
absorption processes within that cell.

\subsection{Thermochemistry}
\label{sec:thermochemistry}

In each cell of the simulation domain, the thermochemical
reaction network is evolved in conjunction with
hydrodynamics, in an operator-splitting manner
(viz. hydrodynamics and thermochemistry are evolved in split
steps, with the same step size in each cycle). A set of
coupled ODEs are solved, reading nominally (note that the
Einstein convention of summation is used),
\begin{equation}
  \label{eq:ode-chem-thermo}
  \begin{split}
    \dfrac{\d n^i}{\d t} & = \mathcal{A}^i_{\;jk} n^j n^k +
    \mathcal{B}^i_{\;j} n^j\ ; \\
    \dfrac{\d \epsilon}{\d t} & = \Gamma - \Lambda\ ;
  \end{split}
\end{equation}
in which the terms involving $\{\mathcal{A}^i_{\;jk}\}$
describe two-body reactions, those in
$\{\mathcal{B}^i_{\;j}\}$ represent photoionization,
photodissociation, and spontaneous decays, and $\Gamma$ and
$\Lambda$ represent the heating and cooling rates per unit
volume, respectively. Stiff as they are, the ODEs of
thermochemistry evolution can be solved with multi-step
implicit method, which, by using the GPUs, is computed at
costs comparable to the hydrodynamics (\method).

As a direct product of planet formation that takes place in
protoplanetary disks (PPDs), the {\it primordial} atmosphere
of planets are expected to involve thermochemical processes
similar to PPDs. We hence inherit the thermochemical network
and recipes from WG17. Hereby we briefly summarize the
thermochemical mechanisms involved and pertinent references:
\begin{itemize}
\item ``Standard'' two-body interactions in the \verb|UMIST|
  database (\citealt{UMIST2013}; note that the photochemical
  reactions therein are not suitable for our radiation
  field, therefore they are excluded).
\item Photoionization of atoms and molecules
  (\citealt{Verner+Yakovlev1995, Verner+etal1996};
  photoionization of carbon atoms for FUV photons are
  subject to cross-shielding, see
  \citealt{1985ApJ...291..722T}).
\item Photodissociation of \chem{H_2} (subject to
  self-shielding, see \citealt{1996ApJ...468..269D}; note
  also that the FUV pumping of \chem{H_2} and subsequent
  reactions are also included, see discussions in
  \citealt{1985ApJ...291..722T}), \chem{CO} (subject to
  self-/cross-shielding, \citealt{2009A&A...503..323V}), and
  \chem{H_2O} \citep{2014ApJ...786..135A}.
\item Dust-assisted molecule formation
  \citep{Bai+Goodman2009,2014ApJ...786..135A} and
  recombination (\citealt{1987ApJ...320..803D,
    2001ApJS..134..263W}; see also the compilation in
  \citealt{2006A&A...445..205I}).
\item Photoelectric effects of dusts
  \citep{2001ApJ...554..778L,2001ApJS..134..263W}.
\item Dust-gas heat accommodation
  \citep{2001ApJ...557..736G, DraineBook}.
\item Atomic cooling (\citealt{1985ApJ...291..722T}; for
  escape probability see \citealt{1981ApJ...250..478K}).
\item Ro-vibrational cooling of molecules
  \citep{1993ApJ...418..263N, 2010ApJ...722.1793O}.
\end{itemize}

\section{Fiducial model}
\label{sec:fiducial-profile}

\begin{figure*}
  \centering
  \includegraphics [width=7.1in, keepaspectratio]
  {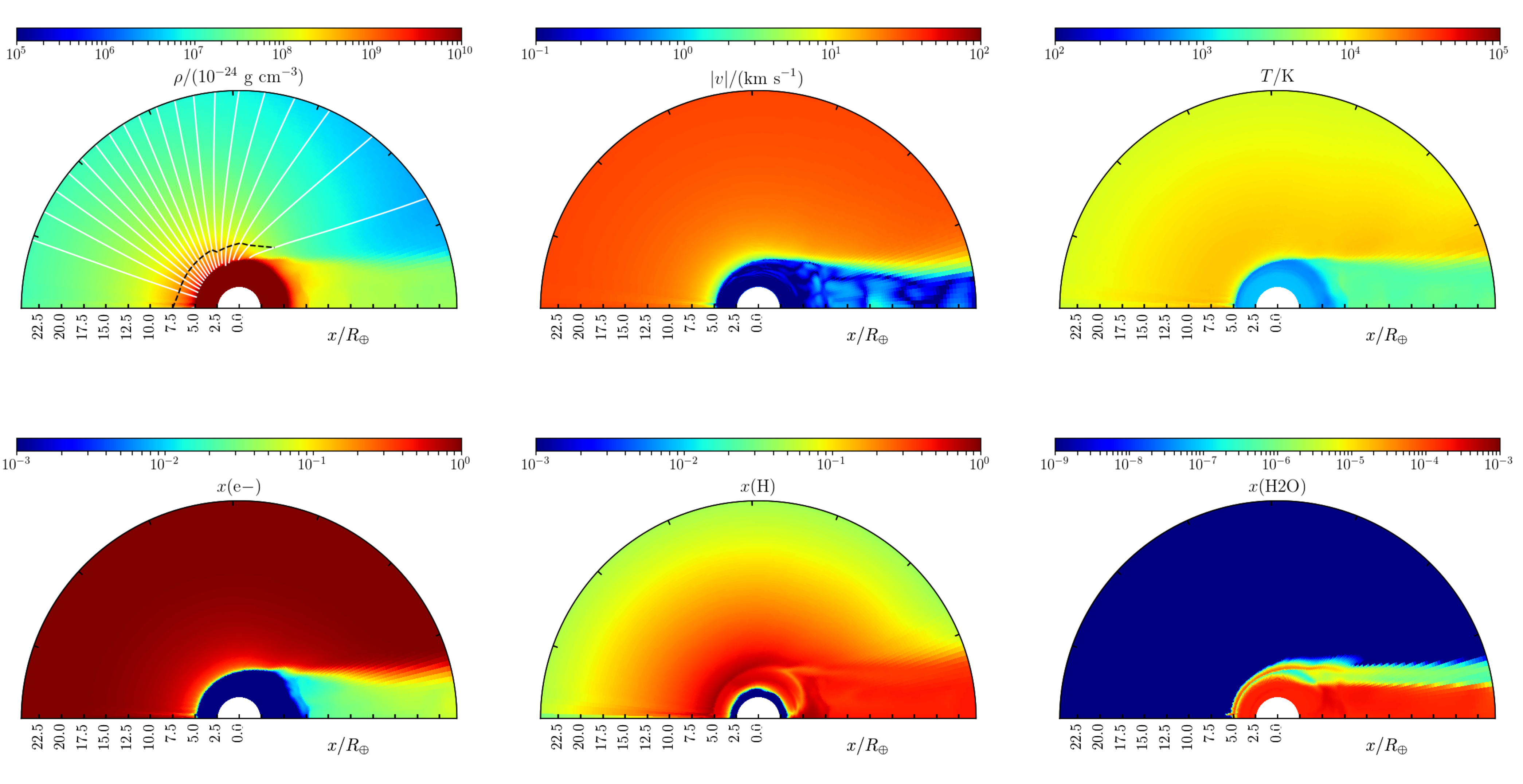}
  \caption{Meridional slice plot of the fiducial model at
    steady state, averaged through the final $0.05~\yr$ of
    the fiducial simulation. The top row presents fluid
    properties, including mass density $\rho$ (upper left
    panel; colormap saturated for
    $\rho>10^{-10}~\g~\cm^{-3}$ to present the wind better),
    magnitude of velocity $|v|$ (upper middle), and
    temperature $T$ (upper right). The bottowm row exhibits
    the relative abundances of important chemical species,
    including electons (lower left), neutral hydrogen (lower
    middle), and water (lower right). The streamlines (white
    solid curves; separated by
    $2\times 10^{-11}~M_\oplus~\yr^{-1}$ mass loss rate) and
    transonic surface (black dashed curve) are overlaid in
    the density panel. Regarding the way of presentation in
    this figure, rays for radiative transfer are injected
    from the left, and all rays travel horizontally across
    the simulation domain. Note the hydrostatic ``tail'' on
    the night side over the planet; see
    \S\ref{sec:flow-struct}. }
  \label{fig:slice_fiducial}
\end{figure*}

\begin{figure}
  \centering
  \includegraphics [width=3.3in, keepaspectratio]
  {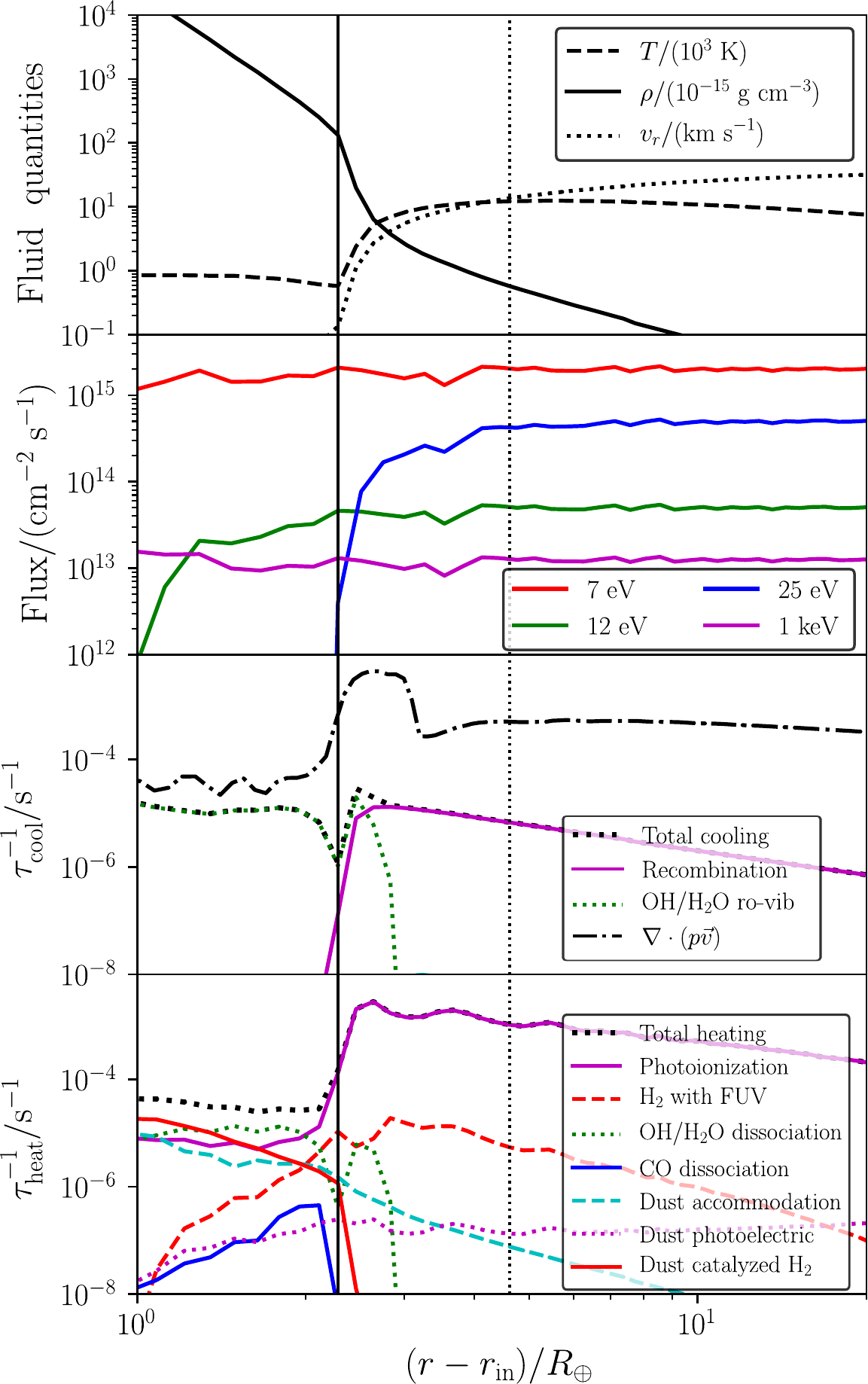}
  \caption{Radial profiles at $\theta = 0.4$ of the
    time-averaged fiducial model. Fluid mechanics profiles
    ($\rho$, $v_r$ and $T$, upper left panel), radiation
    flux in four energy bins (lower left), inverse cooling
    timescale ($\tau^{-1}_\mathrm{cool}$, upper right), and
    inverse heating timescale ($\tau^{-1}_\mathrm{heat}$,
    lower right) in four panels, respectively. Denotation of
    curve shapes and colors are indicated in each individual
    panel. Vertical solid lines show the wind base, while
    vertical dotted lines indicate the sonic point. }
  \label{fig:radial_profile}
\end{figure}

In this section we describe the setup and results of the
fiducial model, which is the reference point of all further
numerical explorations. The main properties of our fiducial
model is summarized in Table \ref{table:fiducial-model}.

\subsection{Fiducial model setup}
\label{sec:fiducial-model-setup}

\begin{deluxetable}{lr}
  \tablecolumns{2} 
  \tabletypesize{\scriptsize}
  \tablewidth{0pt}
  \tablecaption{Properties of the fiducial model}   
  \tablehead{
    \colhead{Item} &
    \colhead{Value}
  }
  \startdata
  Radial domain & $r_\mathrm{in} \le r \le \ 10~r_\mathrm{in}$\\
  & ($r_\mathrm{in}\simeq 2.448~R_\oplus$)\\
  Latitudinal domain & $0\le\theta\le\pi$ \\
  Resolution & $N_{\log r} = 128$, $N_\theta= 128$ \\
  \\
  Planet core & \\
  $M_c$ & $5~M_\oplus$ \\[2pt]
  $R_c$ & $1.495~R_\oplus$ \\[2pt]
  \\
  Planet Atmosphere & \\
  $M_\ad$ & $10^{-2}~M_c$ \\[2pt]
  $T_\eq$ & $886~\K$ \\[2pt]
  $\tau_\kh$ & $10^8~\yr$ \\[2pt] 
  \\
  Radiation flux [photon~$\cm~^{-2}~\s^{-1}$] & \\
  $7~\eV$ (Soft FUV)   & $2\times 10^{15}$ \\
  $12~\eV$ (LW)   & $5\times 10^{13}$ \\
  $25~\eV$ (EUV)  & $5\times 10^{14}$ \\
  $1~\keV$ (X-ray)  & $1.3\times 10^{13}$ \\  
  \\
  Initial abundances [$n_{\chem{X}}/n_{\chem{H}}$] & \\[5pt]
  \chem{H_2} & 0.5\\
  He & 0.1\\
  \chem{H_2O} & $1.8 \times 10^{-4}$\\
  CO & $1.4 \times 10^{-4}$\\
  S  & $2.8 \times 10^{-5}$\\
  Si & $1.7 \times 10^{-6}$\\
  Fe & $1.7 \times 10^{-7}$\\
  Gr & $1.0 \times 10^{-7}$ \\
  \\
  Dust/PAH properties & \\
  $r_\dust$ & $5~\ang$ \\
  $\rho_\dust$ & $2.25~\g~\cm^{-3}$ \\
  $m_\dust/m_\mathrm{gas}$ & $7\times 10^{-5}$ \\
  $\sigma_\dust/\chem{H}$ & $8\times 10^{22}~\cm^2$
  \enddata
  \label{table:fiducial-model}
\end{deluxetable}

We setup the simulation in an axisymmetric spherical polar
grid; dependence on the azimuthal coordinate ($\phi$) is
ignored. The symmetry axis points to the host star, from
which the radiation comes as parallel rays. The grid extends
from $r_\mathrm{in}$ to $10~r_\mathrm{in}$ in radius ($r$)
and $0$ to $\pi$ in co-latitude ($\theta$).  $r_\mathrm{in}$
is defined as the radius at which the static density equals
to a reference value $\rho_\mathrm{in}$, which can be
smaller than $\rho_\rcb$ since photoevaporation only affects
the outermost part of isothermal layer.  In this paper we
choose $\rho_\mathrm{in} = 10^{-7}~\g~\cm^{-3}$ unless
specifically noted. For the fiducial model,
$r_\mathrm{in} = 2.448~R_\oplus$, which guarantees that all
relevant dynamical, radiative and thermochemical processes
are taking places inside the simulated domain.\footnote{For
  the fiducial model, we vary the value of
  $\rho_\mathrm{in}$ by $\pm 1$ orders of magnitude for
  different $r_\mathrm{\in}$, to confirm that the mass loss
  rate is invariant up to $\sim 5\%$.} Outflow boundary
conditions with a radial flow limiter are imposed at
$r=10~r_\mathrm{in}$, and reflecting boundary conditions at
$r=r_\mathrm{in}$, while $\theta = 0$ and $\theta = \pi$ are
polar boundaries. The resolution is $128$ radial by $128$
latitudinal; the radial zones being logarithmically spaced,
and the latitudinal zones equally spaced. Rays of high
energy radiation are injected at the $r=10~r_\mathrm{in}$
boundary in the $\theta$ range $0<\theta < \pi / 2$,
carrying uniform flux density in each energy bin. All rays
are parallel to the symmetric axis.

The fiducial model has a $M_c=5~M_\oplus$ rocky core, whose
radius is therefore $R_c\simeq 1.495~R_\oplus$ (see
\S\ref{sec:atmosphere}). The gravitational field is set
according to the core; self-gravity of the atmosphere is
ignored. Outside the core, we set an adiabatic envelope with
$M_\ad = 10^{-2}~M_c$, surrounded by an isothermal layer
characterized by $T_\eq = 886~\K$ (eq. \ref{eq:T_eq}). The
initial density profile is set according to the discussions
in \S\ref{sec:atmosphere}, with $\tau_\kh = 10^8~\yr$ (see
eq. \ref{eq:tau-kh}). This fiducial model has
$\beta_\ad = 3.94$ and
$\rho_\rcb = 1.1\times 10^{-2}~\g~\cm^{-3}$. The specific
entropy in the adiabatic atmosphere is $\sim 8.3~\kb$ per
baryon.

Host star luminosities in high energy photon are calibrated
as $L_\euv + L_\xray = 10^{30}~\erg~\s^{-1}$
(e.g. \citealt{2012MNRAS.425.2931O}; OW17). In each energy
bin the luminosity is set according to the
$t < 0.1~ \mathrm{Gyr}$ SED concluded in
\citet{2005ApJ...622..680R}, which approximately reads
$L(7~\eV) = L(25~\eV) = L(1~\keV) = 0.5\times
10^{30}~\erg~\s^{-1}$, and
$L(12~\eV) = 0.5 \times 10^{29}~\erg~\s^{-1}$. Those
luminositis are converted into fluxes at $a=0.1~\au$,
assuming that the stellar radiation is isotropic. The
initial abundance of chemicals is the same as WG17, which is
a subset of \citet{2008ApJ...683..287G}, defined by the
values in Table \ref{table:fiducial-model} (where
$n_{\chem{H}}$ is the number density of hydrogen nuclei).

Dust (including PAH) is one of the most important mechanisms
that maintain the temperature in the isothermal region: in
optical and infrared bands where the radiation from the
central star is most energetic, dusts provide most of the
opacity. The abundances of PAH in exoplanet atmospheres,
however, are still unconstrained due to difficulties in
observation. Observations within the solar system suggest
that relatively high concentration of PAH is possible
(e.g. \citealt{2013ApJ...770..132L} shows that the mass
fraction in PAH is $\sim 2\times 10^{-3}$ in the fully
evolved nitrogen-rich atmosphere of Titan, with $\sim 34$
carbon atoms per PAH particle on average). Similar to WG17,
we use the PAH at $5~\ang$ as a proxy of all dusts, with
abundance $10^{-7}$ per hydrogen atom. The dust-to-gas mass
ratio is then $0.7\times 10^{-4}$, and
$\sigma_{\rm dust}/\chem{H}=8\times10^{22}~\cm^2$ for the
geometric cross section. Instead of calculating the
radiative transfer of optical and infrared radiation, we set
the dust temperature $T_\dust = T_\eq$ everywhere in the
simulation for simplicity. The emission power per dust
surface area is proportional to approximately the sixth
power of dust temperature considering dust emissivity
\citep[see e.g.][] {DraineBook}; slight deviation of
$T_\dust$ from $T_\eq$ will result in rapid restoration of
$T_\dust$ back to $T_\eq$.

The fiducial model is run through $\sim 10^{-1}~\yr$ to
guarantee that the system reaches quasi-steady state,
especially that the photospheres of high energy radiation in
four bins do not move. We confirm that the system is already
steady after only $\sim 5 \times 10^{-3}~\yr$: the dynamical
timescale across the simulation domain is at the order of
$t_\mathrm{dyn} \sim 40~R_\oplus / v_r\sim 10^{-4}~\yr$, and
the quasi-steady state is established after only a few
$t_\dyn$.

\subsection{Fiducial model results}
\label{sec:fiducal-results}

The meridional plots, showing the structure of the fiducial
model in the quasi-steady state, are displayed in Figure
\ref{fig:slice_fiducial}. The flow structure is shown by
white streamlines overlaid on the mass density panel (upper
left), separated by constant mass loss rate
$2\times 10^{-11}~M_\oplus~\yr^{-1}$ (integrated through the
polar and azimuthal region between neighbor
streamlines). Streamlines are only plotted in regions with
positive ``Bernoulli parameter'', defined as
\begin{equation}
  \label{eq:bernouli}
  \mathcal{B} \equiv \dfrac{v^2}{2} +
  \dfrac{\gamma p}{(\gamma-1)\rho} + \Phi\ ,
\end{equation}
where $v$ is the magnitude of velocity vector, and $\Phi$ is
the gravitational potential. The surface where the
streamlines terminate is considered as the base of
photoevaporation outflow, which is located at
$r\simeq 5~R_\oplus$. In Table \ref{table:fiducial-radii} we
present several radii as characteristic locations of
different physical processes. Note that $R_\euv$ is the
radius of EUV photosphere, defined as where $F_\euv$ drops
to $10^{-2}$ of the unattenuated value, which also defines
the wind base in our following analyses. \footnote{by
  numerical experiments, we find that this criteria result
  in an almost invariant $\rho$ at $R_\euv$ for different
  physical parameters.}
\begin{deluxetable}{lr}[t]
\tablecolumns{2}
\tabletypesize{\scriptsize}
\tablewidth{200pt}
\tablecaption{Characteristic radii of the fiducial model}
\tablehead{
  \colhead{Item} &
  \colhead{Value}
}
\startdata
$R_c/R_\oplus$ & 1.50 \\
$r_\rcb/R_\oplus$ & 1.91 \\
$r_{\mathrm{in}}/R_\oplus$ & 2.45 \\
$R_\euv/R_\oplus$ & 4.78 \\
$R_s/R_\oplus$ & 7.24
\enddata
\tablecomments
{The fiducial model is defined in 
Table \ref{table:fiducial-model}.}
\label{table:fiducial-radii}
\end{deluxetable}

\subsubsection{Radial profiles of thermodynamics}
\label{sec:radial-prof}

Along a typical radial column at co-latitude $\theta = 0.4$
(where the streamline is almost radial), the hydrodynamical
and microphysical profiles are shown in Figure
\ref{fig:radial_profile}. Only important mechanisms of
heating and cooling are included in the lower two panels,
where the inverses of heating/cooling timescales are defined
as internal energy density divided by the cooling/heating
rate,
$\tau^{-1}_\mathrm{cool,heat} \equiv \epsilon /
f(\mathrm{cool,~heat})$ ($\epsilon$ is the interal energy
density of the gas; not to be confused with the
Kelvin-Helmholtz timescale in \ref{sec:isothermal-reg}).
The term $\nabla\cdot(p\vec{v})$, as another kind of
``cooling'', consists of adiabatic expansion and radial
acceleration and characterizes the rate at which thermal
energy is converted into kinetic energy. The heating curve
marked by ``\chem{H_2} with FUV'' includes photodissociation
and FUV pumping processes by LW FUV photons (see also
\S\ref{sec:thermochemistry}).

Below the wind base, there is a small dip in temperature,
thanks to dramatic expansion as the gas is
accelerated. Inside the isothermal layer where EUV photons
cannot reach, $|\nabla \cdot (p\vec{v})|$ of the gas
creeping outwards at relatively very high density and very
slow speed consumes the majority of injected energy, at a
rate about three times the dissipation by ro-vibrational
cooling of \chem{H_2O} and \chem{OH}. Spatial locations of
the peaks in cooling rate by \chem{H_2O}/\chem{OH} agree
with the spatial distribution of water molecules (see Figure
\ref{fig:slice_fiducial}). Note that there is a layer of
\chem{H_2O} detached to the bulky molecular atmosphere,
where the re-formation rate of \chem{H_2O} exceeds the
photodissociation and thermal collisional destruction
rate. Cooling mechanisms via recombination and \lya, in
contrast, are negligible. On the heating side,
photoionization brought by X-ray and LW photons,
photodissociation of \chem{H_2O} and \chem{OH}, re-formation
of \chem{H_2} on dust surface,
and photoelectric effect of dust grains, are the four
comparable mechanisms that are major heating sources.

From the profiles presented by Figure
\ref{fig:slice_fiducial}, we observe that the day hemisphere
has a clear outflow above the isothermal layer. The outflow
becomes supersonic at the black dashed curve (marking the
sonic points), which confirms that it is a wind rather than
a ``breeze''. As is seen in Figure \ref{fig:radial_profile},
in the subsonic part of wind, the thermodynamics of gas is
dominated by photoionization and
$\nabla\cdot(p\vec{v})$. Beyond the sonic point, radial
acceleration almost vanishes, and $|\nabla\cdot(p\vec{v})|$
gradually surpasses photoionization heating, which causes a
slight ddecreasedrop in temperature.

\subsubsection{Neutral hydrogen and planet size in \lya}
\label{sec:lya-photosph}

We notice that neutral hydrogen atmos still exist at a
considerable fraction in the wind, thanks to the capability
of dealing with non-equilibrium thermochemical
processes\footnote{We confirm that the existence of neutral
  hydrogen in the wind does not affect the structure of
  photoevaporating atmosphere by a test run, whose outer
  boundary is set at $r=10^2~r_\mathrm{in}$ and the radial
  number of zones is doubled so that the mesh configuration
  in the innermost $10~r_\mathrm{in}$ is identical to the
  fiducial run.}. The timescale of hydrogen photoionization
is roughly $(F_\euv \sigma)^{-1} \sim 0.5~\hr$, allowing a
fluid element in the wind to travel $\sim 8.4~R_\oplus$
before its fraction of neutral hydrogen decreases by one
e-fold. The low ionization fraction prohibits the
propagation of $\lya$ into the isothermal layer: the
line-center optical depth is $\sim 10^{6-7}$, which allows
us to ignore \lya radiative transfer safely.  Meanwhile,
neutral hydrogen in the wind makes the observed size of
planet in \lya much bigger than other bands. The
dimensionless equivalent width on a wide \lya profile
denoted by $\phi(u)$, is estimated as (here $u$ is the
line-of-sight velocity),
\begin{equation}
  \label{eq:lya-eq-width}
  W = \left[\int \d u ~\phi(u)\right]^{-1} \times
  \left[\int \d u~\phi(u) (1 - \e^{-\tau(u)})\right]\ .
\end{equation}
We assume that $\phi(u)$ is a Gaussian profile with FWHM
$\sim 200~\km~\s^{-1}$, and that $\tau(u)$ is another
Gaussian with FWHM $\sim 50~\km~\s^{-1}$, combining the bulk
and thermal motion of the outflow. By integrating $\tau$ and
then $W$ along different lines of sight, we find that
$W\sim (1-\e^{-1})$ at an impact parameter
$b \sim 11~R_\oplus$, which roughly indicates the observed
size in \lya during transits. The excess of obsereved size
of planets in \lya compared to optical is readily observed
in Jupiter-size objects \citep[e.g.][]{2010A&A...514A..72L}.
According to our simulations, the same holds true for
sub-Neptune planets.

\subsubsection{Flow structures and the mass loss rate}
\label{sec:flow-struct}

Above the night hemisphere, where no high energy photons can
ever reach, there exists a roughly isothermal ``tail'' at
$\rho \sim 10^{-15}-10^{-16}~\g~\cm^{-3}$ and
$T\sim 2\times 10^3~\K$. The tail is almost static; the
velocity magnitude is roughly below
$10^{-1}~\km~\s^{-1}$. The streamline structures and the
evolution history both reveal that the tail is brought about
by the gas flowing along the streamline originating from
$\theta \simeq \pi/2$, similar to the Bernoulli effect for
isenthalpic flows. Existing works of planet atmosphere
suggest much larger tails (outside the Roche lobe of the
planet) which are typically attributed to ram pressure of
stellar wind or radiation pressure of the central star
\citep[e.g.][] {2016ApJ...820....3C,2016A&A...591A.121B},
while the tail here comes from the thermal wind of the
planet itself. The fate of this ``smaller'' tail, however,
has to be studied with full three-dimensional simulations in
order to include the effects of planet orbital motion, which
we will tackle in an upcoming paper.

By integrating the mass flux through the outer boundary of
$r$, we obtain a mass loss rate
$\dot{M}\simeq 4\times 10^{-10}~M_\oplus~\yr^{-1}$. For the
fiducial case alone, multidimensionality does not contribute
to the mass loss rate appreciably. Multiplying the radial
mass flux at $\theta = 0$ (at the substellar point) by
$2\pi$ (the solid angle of day hemisphere), the estimated
mass loss rate is only $\sim 3~\%$ bigger than the value
measured from simulation. The dimensional effects (i.e. 2.5
dimensions with axisymmetry) are nonetheless manifested by
models with strong stellar wind ram pressure (see
\S\ref{sec:stellar-wind-ram}).

 
\section{Exploring the parameter space}
\label{sec:explore-param}

\begin{figure}
  \centering
  \includegraphics [width=2.8in, keepaspectratio]
  {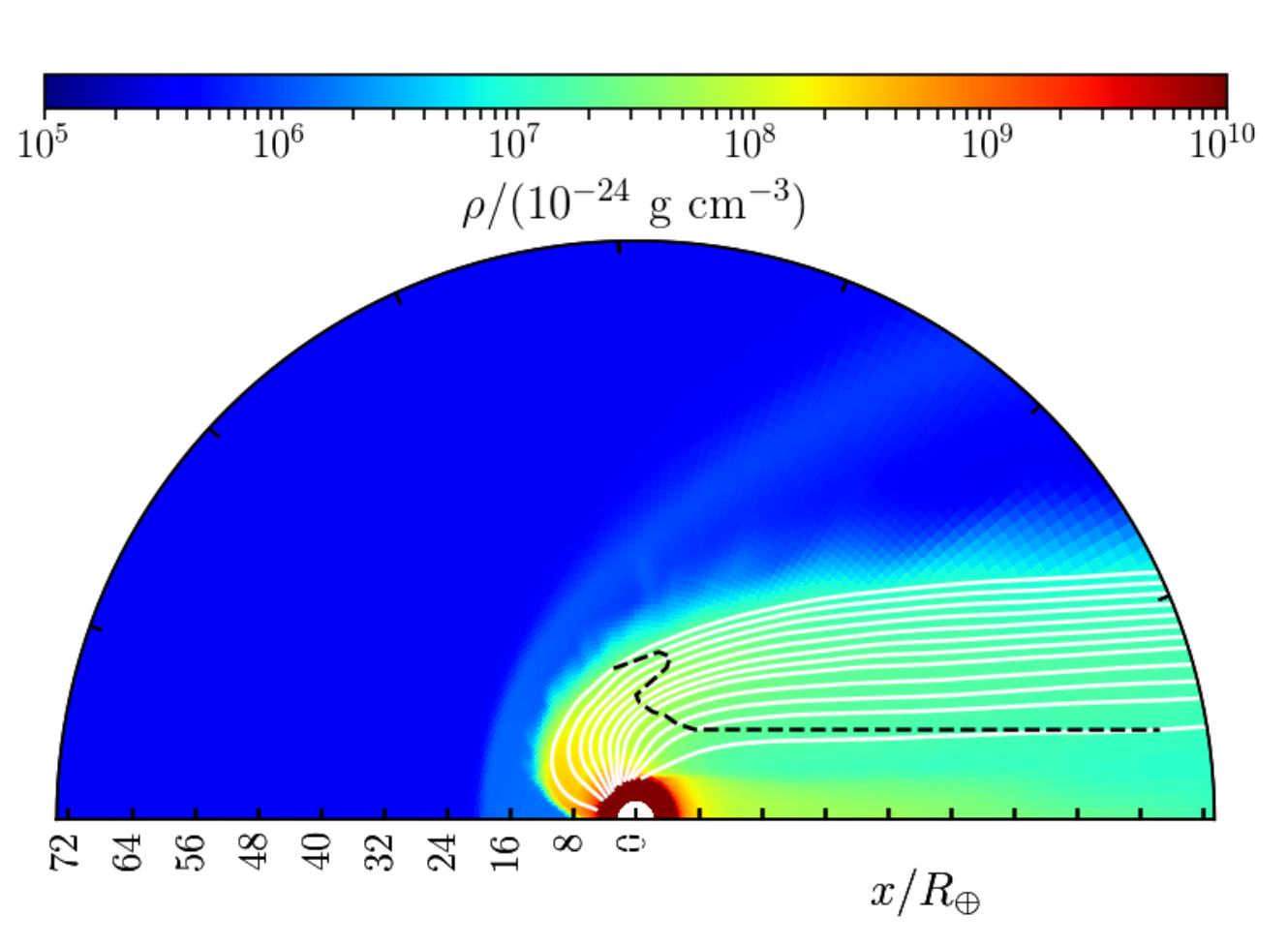}
  \caption{Meridional slice plot of the
    $p_\ram = 10^{-3}~\dyn~\cm^{-2}$ case in Models W, where
    a strong stellar wind ram pressure is included. The
    colormap exhibits the mass density $\rho$, showing the
    time-averaged results for the final $10^{-2}~\yr$. The
    denotation of overlays in this figure are identical to
    the upper left panel of Figure
    \ref{fig:slice_fiducial}. See
    \S\ref{sec:stellar-wind-ram} for the details of
    simulation setup. \footnote{An animation showing the
      evolution of this model is available as online
      supplement material}. }
  \label{fig:model_w}
\end{figure}

\begin{figure}
  \centering
  \includegraphics[width=3.3in, keepaspectratio]
  {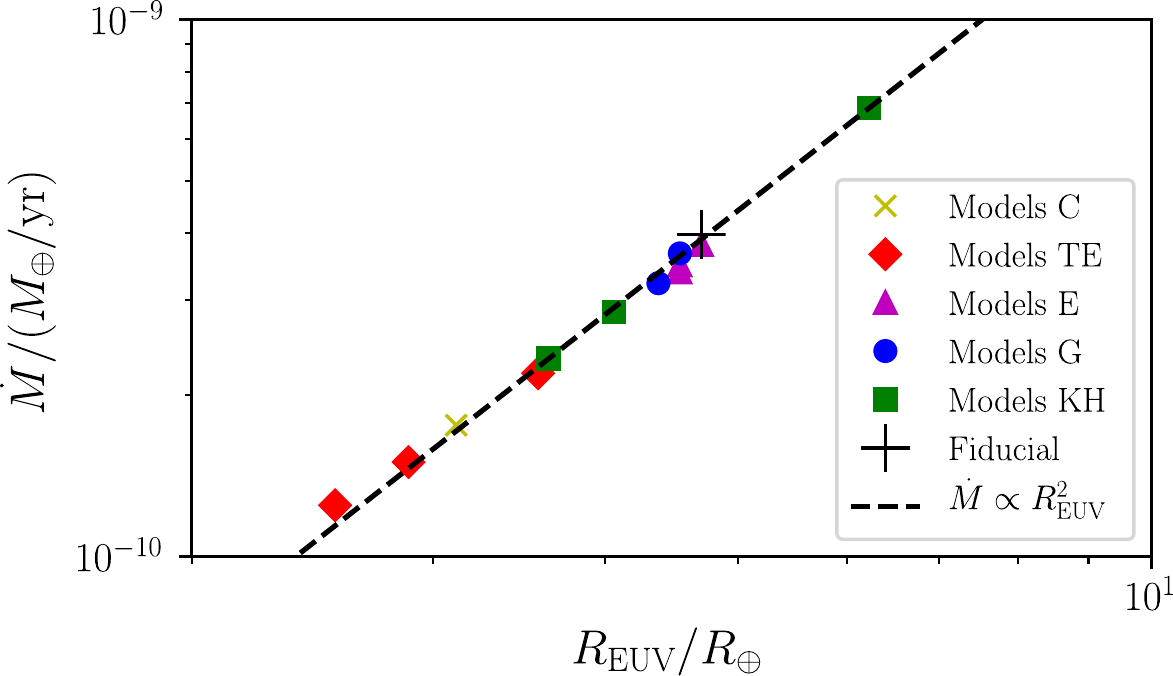}
  \caption{Scattered plots showing the $\dot{M}$-$R_\euv$
    relations for Models E, G, KH, TE, and C, as well as the
    fiducial model. Makers of each model series are
    indicated by the legend. The simple relation
    $\dot{M}\propto R_\euv^2$ is shown by a dashed line.  }
  \label{fig:scat_mdot_reuv}
\end{figure}

\begin{figure*}
  \centering
  \includegraphics [width=6.9in, keepaspectratio]
  {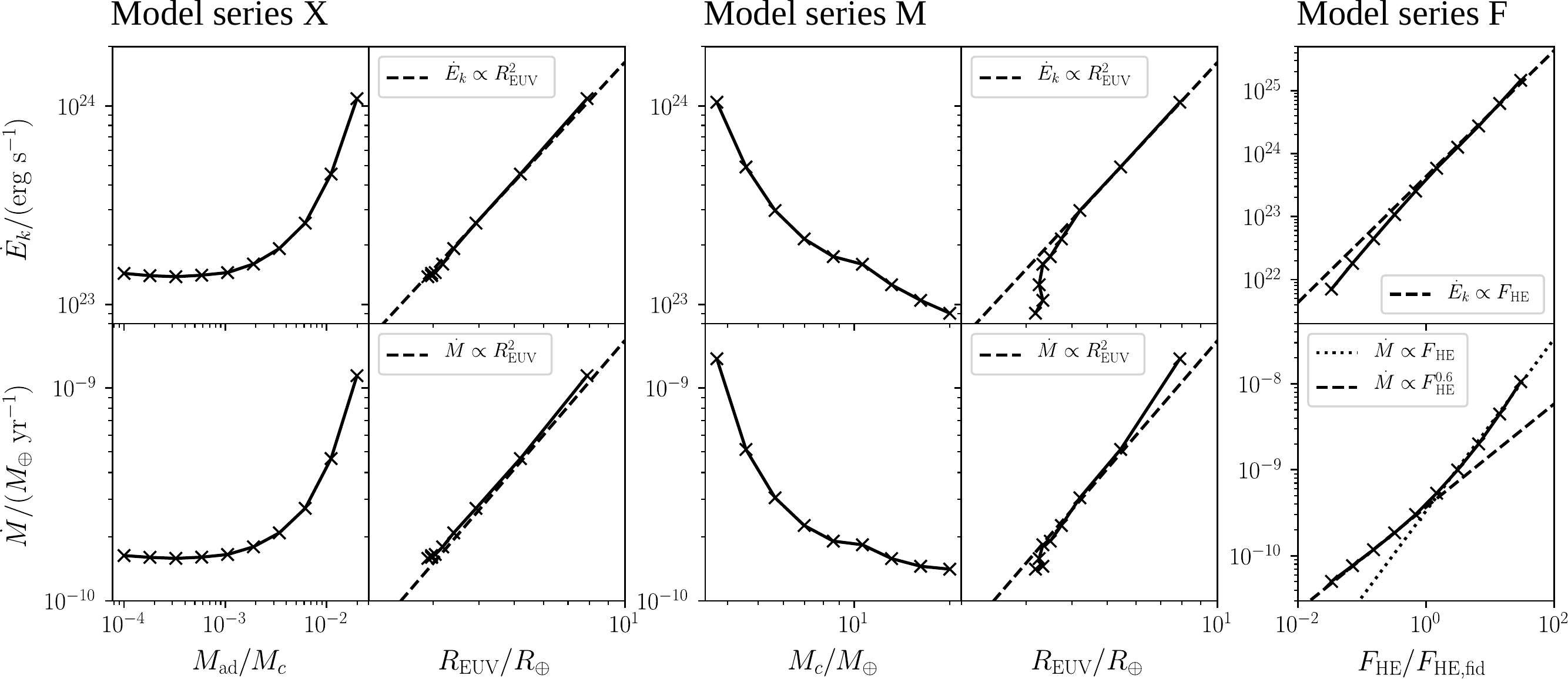}
  \caption{Simulation results of model serieses X (varying
    envelope mass fraction; left block), M (varying core
    mass; central block), and $F$ (varying high energy flux,
    right block). Each data point, representing a model in
    its quasi-steady state, is indicted by a cross
    (``$\times$''). In some panels, dashed or dotted lines
    present simple power-laws (indicated by legends in the
    panels) to help the reader recognize the general trend
    of variation. Panels related to $R_\euv$ are not shown
    for model series F as $R_\euv$ does not vary appreciably
    in the series. $y$-axes of the panels on the top-row all
    present $\dot{E}_k$ (the kinetic energy carried by the
    outflow, in $\erg~\s^{-1}$), while those of the
    bottom-row panels present $\dot{M}$ (the mass loss rate,
    in $M_\oplus~\yr^{-1}$). }
  \label{fig:model_series}
\end{figure*}

\begin{deluxetable}{ccc}[t]
\tablecolumns{3}
\tabletypesize{\scriptsize}
\tablewidth{240pt}
\tablecaption{Models exploring the parameter space}
\tablehead{
  \colhead{Model} &
  \colhead{Variable} &
  \colhead{Values} 
  \\
  \colhead{Series} &
  \colhead{} &
  \colhead{} 
  \\
  \colhead{(1)} &
  \colhead{(2)} &
  \colhead{(3)} 
}
\startdata
& Microphysics parameters & \\
E & Energy bin turned off (eV) & $\{7,\ 12,\ 25,\ 10^3\}$ \\
G & Dust grain abundance & $\{10^{-8},\ 10^{-9}\}$ \\
\\
& Atmosphere parameters & \\ 
KH & $\tau_\kh/(10^8~\yr)$ & $\{0.3,\ 3,\ 10\}$ \\
TE & $T_\eq/\K$ & $\{411,\ 518, 703\}$ \\
X & $M_\ad/M_c$ & $[10^{-4}, 0.02]$ \\
\\
& Planet and stellar properties & \\
W$^\dagger$ & $p_\ram/(\dyn~\cm^{-2})$ & $\{10^{-4},\
10^{-3},\ 10^{-2} \}$ \\  
C & Core composition & $\{\mathrm{Iron,\ Water}\}$ \\
M & $M_c$ & $[3, 20]$  \\
F & Relative high energy flux & $[1/30,30]$
\enddata
\tablecomments
{
  (1) Model series identifier.
  (2) Description of the varied quantity in the model
  series.
  (3) Sets of values of the variable. Squared brackets
  $[l,u]$ denote that the variable will take several
  different values between lower limit $l$ and upper limit
  $u$, while braces ``$\{\}$'' indicate that the variable
  will select one of the discrete values in the braces at a
  ime. \\
  $\dagger$: See \S\ref{sec:stellar-wind-ram} for details of
  simulation setup. }
\label{table:parameter-explore}
\end{deluxetable}

We have run serieses of simulations to explore the effects
of different parameters. To manifest the impact of each
parameter more clearly, each run in this subsection differs
from the fiducial model by only one parameter unless
specifically stated. Table \ref{table:parameter-explore}
lists the model serieses [indicated by ``Model(s) $Y$'' for
the model series $Y$ in what follows] with brief
descriptions.

\subsection{Stellar wind ram pressure and the effects of
  multidimensions}
\label{sec:stellar-wind-ram}

For a typical radial column above the day hemisphere, in
which the flow is virtually radial, we derive the total
pressure of outflow (including thermal pressure and fluid ram
pressure), from conservation of momentum (here
$v_{r,\infty}$ is the terminal radial velocity of outflows),
\begin{equation}
  \label{eq:p-euv}
  \begin{split}
    p_{\tot} &
    \simeq 10^{-3}~\dyn~\cm^{-2}\times
    \left(\dfrac{\dot{M}}{10^{-10}~M_\oplus~\yr^{-1}}\right)
    \\
    & \times
    \left(\dfrac{v_{r,\infty}}{34~\km~\s^{-1}}\right)
    \left(\dfrac{r}{5~R_\oplus}\right)^{-2}\ .
  \end{split}
\end{equation}
This pressure is typically much greater than ambient
pressure (eq. \ref{eq:press-wind}), which allows us to
safely ignore the ambient constraints once photoevaporation
outflow is launched. If the model is 1D and spherical
symmetric, when the ram pressure exerted by the stellar wind
$p_\ram$ is comparable to the total pressure near the sonic
point (namely $p_{\tot,s}$), one would expect that the
supersonic outflow is quenched while only a subsonic breeze
is possible (see also M-CCM09).

However, considering the multidimensional reality, it is
more likely that the supersonic planetary winds would divert
to the night hemisphere instead of being totally chocked. We
hence compute simulation Models W, in which we setup
significant inflow at the outer $r$-boundary, in the range
$0<\theta < \pi/2$ (the boundary condition is the same as
fiducial in $\pi/2 < \theta < \pi$). The inflow is parallel
to the radiation fluxes, at velocity $v = 500~\kms$, and
temperature $T = 10^5~\K$. Mass density of the inflow is the
variable that controls $p_\ram$ in Models W:
$\rho = 4\times 10^{-19}~\g~\cm^{-3}\times
(p_\ram/10^{-3}~\dyn~\cm^{-2})$.  Three different ram
pressures are tested respectively,
$(p_\ram / \dyn~\cm^{-2}) \in \{10^{-4},\ 10^{-3},\
10^{-2}\}$, compared to the total outflow pressure at the
sonic point
$p_{\tot,\s}\simeq 1.8\times 10^{-3}~\dyn~\cm^{-2}$ for the
fiducial model. The outer boundary is $30~r_\mathrm{in}$ so
that all important hydrodynamic features are correctly
included in the simulation domain, and the number of radial
zons is adjusted accordingly in order to keep the resolution
of the innermost region identical to the fiducial one.

In their quasi-steady states, all cases in Models W still
have significant supersonic outflows originating from the
day hemisphere flowing to the night side.  Although the flow
morphology changed dramatically, the mass loss rate is
comparable to the fiducial model.  Previous one-dimensional
calculations carried out by M-CCM09 could not capture the
multidimensional effects, thus incorrectly predicted the
quenching of supersonic photoevaporative outflow when ram
pressure becomes significant. For each individual model:
\begin{itemize}
\item The $p_\ram = 10^{-3}~\dyn~\cm^{-2}$ case has mass
  loss rate $\dot{M}\simeq (3.0\pm 1.1)~M_\oplus~\yr^{-1}$
  (the uncertainty here is the standard deviation over the
  final $10^{-2}~\yr$, reflecting the variation amplitude of
  the outflow).  In Figure \ref{fig:model_w}, we present a
  meridional plot of mass density with streamlines overlaid
  for this case, based on the time-averaged results during
  the final quasi-steady state ($10^{-2}~\yr$) of the
  simulation.  Although the outflows are suppressed by the
  post-shock external wind near the substellar point, they
  still find their ways out on the night side and become
  supersonic. The discontinuity surface is rather turbulent
  thanks to Kelvin-Helmholtz instability (as the fluid speed
  behind the bow shock is still $\sim 10^2~\kms$), which in
  turn affects the shape of the shock as the post-shock flow
  is subsonic. Therefore we do not observe sharp transitions
  for the bow shock and the discontinuity in Figure
  \ref{fig:model_w}.
\item The $p_\ram = 10^{-4}~\dyn~\cm^{-2}$ case also has
  turbulent supersonic outflow, with mass loss rate
  $\dot{M}\simeq (3.3\pm 1.0)~M_\oplus~\yr^{-1}$.
\item The test case $p_\ram = 10^{-2}~\dyn~\cm^{-2}$ still
  has $\dot{M}\simeq (3.1\pm 0.1)~M_\oplus~\yr^{-1}$ mass
  loss rate, even when the ram pressure is one order of
  magnitude greater than the fiducial total pressure at the
  sonic surface $p_{\tot,s}$. The supersonic outflow becomes
  fairly laminar in this case.
\end{itemize}

\subsection{Different bands of radiation}
\label{sec:radiation-bands}

Models E, varying the SED of incident high energy radiation
by turning off one band of radiation at a time, help us to
understand in which band is the high energy photons most
relevant to photoevaporation. Simulation results reveal that
only the EUV photons have primary impact on the mass loss
rate. By turning off the EUV flux, the mass loss rate drops
to $\dot{M}\lesssim 10^{-12}~M_\oplus~\yr^{-1}$, while the
gas is creeping outwards at a radial velocity
$v_r\lesssim 10^{-1}~\km~\s^{-1}$. This is understood by
comparing the depth of gravitational potential well, which
is roughly $\sim 1~\eV$ per proton for a $5~M_\oplus$ planet
core at $r\sim 3~R_\oplus$, to the energy per particle
deposited by high energy photons.

When an EUV photon is absorbed by a hydrogen atom/molecule,
photoionization processes deposit roughly $10~\eV$ per
reaction of energy to the post-interaction particle. LW
photons, in comparison, deposit-es only $\sim 0.5~\eV$ of
energy into each hydrogen atom by dissociating a \chem{H_2}
molecule \citep[e.g.][]{1979ApJS...41..555H}, which is
marginally sufficient to free it from the potential well. As
EUV photons interact with the most abundant elements
(hydrogen and helium), the energy injected is not
considerably diluted. In contrast, energy injected by soft
FUV and X-ray photons, which interact predominantly with
species at relatively low abundance (especially water and
dust grains), experiences significant dilution. Soft FUV and
X-ray photons penetrate to higher depths where the number
densities of hydrogen nuclei are rather high
($\sim 1\times 10^{14}~\cm^{-3}$ for soft FUV and
$\sim 3\times 10^{14}~\cm^{-3}$ for X-ray). At those high
densities, energy deposited by those photons is easily
transferred to coolants or accommodated by dusts, and then
re-radiated as infrared photons that are not efficient in
heating the gas at all.

Meanwhile, when bands other than EUV are turned off, the
mass loss rate is only secondarily affected: being
responsible to modifying the temperature in quasi-isothermal
layer, turning off a non-EUV band of radiation leads to (a)
slight shrinking of the quasi-isothermal layer and hence EUV
photosphere, and (b) survival of more coolants. As a result,
the mass loss rate decreases by $\sim 5\%$ (LW off) or
$\sim 15\%$ (X-ray or soft FUV off).

\subsection{Configuration of atmospheres}
\label{sec:config-atm}

An important substance maintaining temperature in the
isothermal layer as they are, dust grains (which we use PAH
as the proxy) have nonetheless highly uncertain abundances
(see \S\ref{sec:fiducial-model-setup}). In Models G, we
confirm that the abundance of Gr affects $\dot{M}$ by
competing with the cooling mechanisms in the
quasi-isothermal region and then setup the location and
structure at the EUV photosphere. Reducing the Gr abundance
by one or two orders of magnitude results in a $\sim 10\%$
or $\sim 20\%$ decrease in $\dot{M}$ respectively.

Covering the uncertainties in the hydrostatics of planet
atmosphere, we setup Models KH and TE. $\tau_\kh$ and
$T_\eq$ together characterize the specific entropy and hence
the density profile of the atmosphere (see
\S\ref{sec:isothermal-reg}). Models C, varying the core
density by assuming an iron or water core below the
atmosphere (using the mass-radius relation in
\citealt{2014ApJ...792....1L} for different core
components), also modify the atmospheric density profile
dramatically.

It is worth noting that almost all models discussed above
obey the simple relation, $\dot{M}\proptosim R_\euv^2$, as
$R_\euv^2$ is proportional to the effective area for the
planet to intercept EUV photons.
We exhibit this scaling relatino in Figure
\ref{fig:scat_mdot_reuv}. 

\subsection{Envelope mass fraction and the planet core}
\label{sec:env-mass-frac}

Models X, varying the envelope mass fraction, also clearly
presents the dependence $\dot{M}\propto R_\euv^2$, as is
seen in the left column in Figure
\ref{fig:model_series}. This relation is almost invariant
when the wind starts from different depth in the
gravitational potential well: the kinetic energy at
$v_\wind\simeq 32~\km~\s^{-1}$ overwhelms the gravitational
potential (escape velocity is only $\sim 11~\km~\s^{-1}$ at
$r=5~R_\oplus$ for a $5~M_\oplus$ core) and other energy
balance as a fluid element escapes. The kinetic energy
carried by the outflow, $\dot{E}_k$, also follows the same
proportionality. Thermal energy carried by the outflow is
only $\sim 10^{-1}$ of $\dot{E}_k$, hence does not affect
our discussions.

When the core mass varies in the Models M, the physics
becomes slightly different (see the middle column in Figure
\ref{fig:model_series}). If $M_c \gtrsim 10~M_\oplus$,
$R_\euv$ varies much slower with $M_c$, but the depth of the
potential well starts to become significant. In the
hypothetical case with $M_c= 20~M_\oplus$, $v_{r,\infty}$
drops to $\sim 26~\km~\s^{-1}$, as the gas spent part of its
energy against gravitational potential well when escaping.

\subsection{High energy photon flux and ``cooling
  limited''}
\label{sec:euv-flux-limit}

In Models F (see the right column in Figure
1\ref{fig:model_series}), $R_\euv$ does not vary appreciably
with the luminosity. When the high energy fluxes are weak
($F_\he \lesssim F_{\he,\fid}$), the mass loss rate scales
as $\dot{M}\proptosim F_\he^{0.6}$ (shallower than linear),
while the kinetic energy of outflow drops faster than
linear. Considering the proportionalities
$\dot{M}\propto r^2 \rho v_r$ and
$\dot{E} \propto r^2 \rho v_r^3$, this clearly indicates
that the outflow speed drops drastically at low
luminosities, which in turn increases the efficiency of
converting energy into outflow.  In fact, for
$F_\he/F_{\he,\fid} = 1/30$, we have
$v_{r,\infty}\simeq 12~\km~\s^{-1}$. At high fluxes,
$\dot{M}$ has linear dependence on $F_\he$

Interestingly, M-CCM09 claimed that
$\dot{M}\propto F_\he^{0.6}$ at high $F_\he$ due to
increasing rate of recombination cooling, and that
$\dot{M} \propto F_\he^{0.9}$ otherwise. They concluded that
recombination cooling at high $F_\he$ is the limiting factor
(``recombination limited''). Our models confirmed the power
index $0.6$ when cooling is the major limit. However,
involving detailed thermochemistry reveals that the dominant
coolants are the molecules via ro-vibrational transitions,
while recombination is indeed impossible to remove injected
heat at considerable amounts near the wind base (see
discussions in \S\ref{sec:static-caveat}). At lower $F_\he$,
the abundances of molecular coolants are higher. We hence
summarize the mass loss rate at low $F_\he$ with a more
general term, ``cooling limited'', instead.

\section{Discussions}
\label{sec:discussions}

\subsection{Grid data of photoevaporative model and
  evaporation timescale}
\label{sec:grid-data}

\begin{figure}
  \centering
  \includegraphics[width=3.3in, keepaspectratio]
  {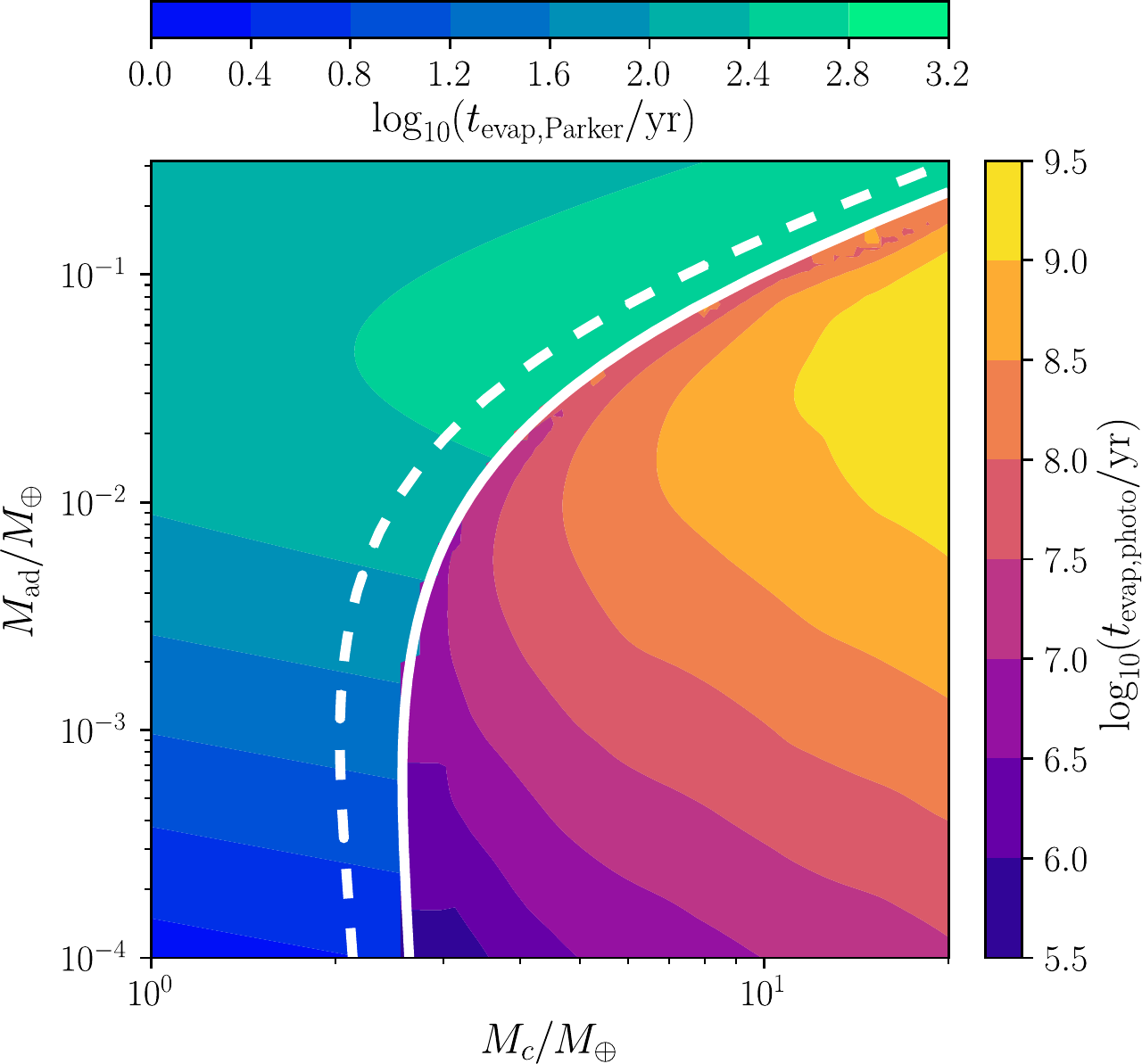}
  \caption{Evaporation timescale of planet atmosphere.  The
    heavy white curve is the critical curve above which the
    isothermal layer pressure cannot be balanced by the
    ambient ($p_\infty< p_\amb$, see \S\ref{sec:mass-loss}),
    and the white dashed curve indicates
    $\rho_\infty \simeq 10^{-13}~\g~\cm^{-3}$, above which
    the pressure balance by outflow is no longer valid (see
    also discussions in \S\ref{sec:grid-data}). $t_\evap$ in
    the region with unbalanced ambient pressure above the
    solid white white curve is calculated with
    eqs. \eqref{eq:massloss-parker} and
    \eqref{eq:massloss-ene}, which should read the
    blue-green color map indicated on the top. $t_\evap$
    based on simulation results, presented by the region
    below the critical curve, should read the purple-yellow
    color map on the right. }
  \label{fig:t_diss_example}
\end{figure}

In order to understand the general trend of $t_\evap$
varying with different planet configurations, we run a grid
of simulations with parameters in the space where
$p_\infty < p_\amb$ (parameters unspecified are identical to
the fiducial model),
\begin{equation}
  \begin{split}
    & \{M_c/M_\oplus,\ M_\ad/M,\
    F_\he/F_{\mathrm{HE,fid}} \}
    \\
    & \in \{[3,20] \otimes [10^{-4},10^{-0.5}]
    \otimes [10^{-2},1]\} \cap \{p_\infty < p_\amb\}\ .
  \end{split}
\end{equation}
The resolution is 10 logarithmically-spaced grid points
along $M_c$, 10 logarithmic along $M_\ad/M_c$, and 3
logarithmic along $F_\he$. After excluding
$p_\infty > p_\amb$ points 261 simulations in total have
been integrated to at least $10^2~t_\dyn$. Figure
\ref{fig:t_diss_example} plots $t_\evap$ the grid at
$F_\he/F_{\mathrm{HE,fid}}=1$, where data in the region
$p_\infty > p_\amb$ (above the heavy white curve) present
the timescale of mass loss through Parker wind
(\S\ref{sec:mass-loss}).

When the envelope of a planet starts evolution in the Parker
wind region, it is expected to be dispersed rapidly, at
timescale shorter than $10^{3}~\yr$, until it becomes a
``bare'' planet ($M_c\lesssim 3~M_\oplus$), or reaches the
photoevaporation region ($M_c\gtrsim 3~M_\oplus$), as we
already discussed in \S\ref{sec:mass-loss}. This scenario is
slightly changed when high energy radiation fluxes
exist. Compared to the ambient, outflows exert much greater
pressure confinement onto the internal static atmosphere at
wind bases (see eq. \ref{eq:p-euv}). However, this kind of
``confinement'' does require the isothermal atmosphere to
allow EUV photons penetrate. Inspecting the location of EUV
photosphere in various models, $R_\euv$ is always found at
radii where $\rho\sim 10^{-13}~\g~\cm^{-3}$ (also see
discussions in M-CCM09); if
$\rho_\infty \gtrsim 10^{-13}~\g~\cm^{-3}$, EUV photons are
unlikely to penetrate, and this mechanism (pressure balance
by outflow) is not valid either. In Figure
\ref{fig:t_diss_example} we present this criterion by a
white dashed curve. The region between two type of pressure
balancing is rather narrow. Therefore, in practice, adopting
the ambient pressure balancing criterion will not
significantly change the story, which is what we will do in
the followings.

Planets in the photoevaporative domain disperse their
envelopes at much longer timescales. In particular, the
timescale peaks at $M_\ad/M_c\sim 10^{-2}$,
$t_\evap \gtrsim 10^8~\yr$ for $M_c\gtrsim 5~M_\oplus $. At
the same $M_c$, planets with $M_\ad/M_c$ above this peak
have rather puffy envelope, and $R_\euv^2$ increases
dramatically faster than $M_\ad$. The power injected by
intercepting EUV photons per $M_\ad$ hence increases,
shortening $t_\evap$ as a result. Below that peak, $R_\euv$
does not shrink appreciably, especially when
$M_\ad/M_c\lesssim 10^{-3}$. $\dot{M}$ is almost constant,
thus $t_\evap\proptosim M_\ad$ becomes shorter and shorter
as a planet loses its envelope mass. Qualitatively those are
similar to the mechanisms proposed in OW17 which are
probably responsible of the bimodal distribution of observed
planet radius. We will discuss this with more details in
\S\ref{sec:evolution}. We particularly noticed that, for
planets with core mass $M_c\lesssim 6~M_\oplus$, evaporation
timescales of their envelopes will be shorter than
$\sim 300~\mathrm{Myr}$ everywhere. In other words, under
fiducial set of parameters, photoevaporation prevents us
from finding rich H/He envelopes on those low mass planets
in evolved systems. This is not a hard limit,
since the photoevaporation conditions vary from system to
system, while $t_\evap$ depends on those conditions rather
sensitively. Nevertheless, observations show the decline in
number of planets with rich atmospheres \citep[see
also][]{2015ApJ...800..135D,2015ApJ...801...41R}. 
We thus suggest that this
decrease can be attributed to photoevaporation.

\subsection{Scaling relations of the mass loss rate}
\label{sec:scaling-mdot}

Based on the simulation grid as well as the explorations in
\S\ref{sec:explore-param}, in almost all cases already
tested, the relation $\dot{M}\propto R_\euv^2$ holds very
well (except for $M_c \gtrsim 10~M_\oplus$, where the depth
of potential well begins to make a difference). We notice
that, in M-CCM09 and OW17, the mass-loss rate is scaled to
an expression proportional to $r_\rcb^3$. The third power
implies two assumptions, if the efficiency is nearly
constant:
\begin{enumerate}
\item Terminal specific energy of the outflow is
  proportional to the depth of gravitational potential well
  at the wind base.
\item The radius of EUV photosphere (approximately the
  same as wind base) is proportional to $r_\rcb$.
\end{enumerate}
Assumption 1 disagrees with our simulations, which indicate
that the specific energy of outflow at infinity is almost
invariant in most cases as it is much greater than the depth
of potential well at wind bases. It is obvious that
assumption 2 does not hold either. By observing our grid of
photoevaporation models, we confirm that in each model
$R_\euv$ is always located at
$\rho\sim 10^{-13}~\g~\cm^{-3}$. Using
eq. \eqref{eq:rho_isothermal}, we estimate by assuming that
EUV photons penetrate to where
$\rho\sim 10^{-13}~\g~\cm^{-3}$,
\begin{equation}
  \label{eq:reuv-estimate}
  R_\euv \simeq r_\rcb \left[ 1 + \tilde{\beta}_\iso^{-1}
    \ln\left( \dfrac{10^{-13}~\g~\cm^{-3}} { \rho_\rcb }
    \right) \right]^{-1}\ ,
\end{equation}
where $r_\rcb$ and $\rho_\rcb$ can be determined
analytically using eqs. \eqref{eq:mass-adiabatic} and
\eqref{eq:m_ad-tau}. Clearly $R_\euv/r_\rcb$ is not
constant. $\tilde{\beta}_\iso$ here should be estimated
using a different $\mu$ from the one used for the adiabatic
layer, as hydrogen molecules in the upper part of the
isothermal layer are partially dissociated by LW and X-ray
photons. We find that $\mu\simeq 1.88~m_p$ fits $R_\euv$ all
models in the simulation grid and \S\ref{sec:explore-param}
with errors $\lesssim 10~\%$ (mostly within $5~\%$).

According to \S\ref{sec:euv-flux-limit}, the dependence of
$\dot{M}$ on incident high energy flux is linear if
$F_\mathrm{EUV}\gtrsim F_\mathrm{EUV,fid}$, and
$\dot{M}\proptosim F_\mathrm{EUV}^{0.6}$
otherwise. Calibrated at the fiducial model, we propose a
semi-empirical formula for mass loss rate that reads,
\begin{equation}
  \label{eq:mdot-fitting}
  \begin{split}
    & \dot{M}\simeq 4.5\times 10^{-10}~M_\oplus~\yr^{-1}
    \times \left(\dfrac{R_\euv}{5~R_\oplus}\right)^2
    \\
    & \quad \times \max\{\mathcal{F},\mathcal{F}^{0.6}\}~
    \min\{1,\mathcal{M}^{-0.5}\}\ ;
    \\
    & \mathcal{F}\equiv
    \dfrac{F_\euv}{F_{\euv,\fid}}\ ;\quad
    \mathcal{M} \equiv \dfrac{M_c}{10~M_\oplus}\ .
  \end{split}
\end{equation}
With $R_\euv$ given by eq. \eqref{eq:reuv-estimate}, we test
the mass loss rate fitting by setting up 100 simulations
with random (but reasonable) combinations of all parameters,
confirming that the error of eq. \eqref{eq:mdot-fitting} is
$\lesssim 20~\%$ for most simulation runs, or
$\lesssim 50~\%$ at worst [most of the worst cases have
relatively large $R_\euv$ ($R_\euv \gtrsim 20~R_\oplus$)
that cannot be accurately estimated by
eq. \eqref{eq:reuv-estimate}].

\subsection{Mass and radius evolution of planet atmosphere}
\label{sec:evolution}

\subsubsection{Evolution of sample planets}
\label{sec:sample-planet-evo}

\begin{figure}
  \centering
  \includegraphics[width=3.3in, keepaspectratio]
  {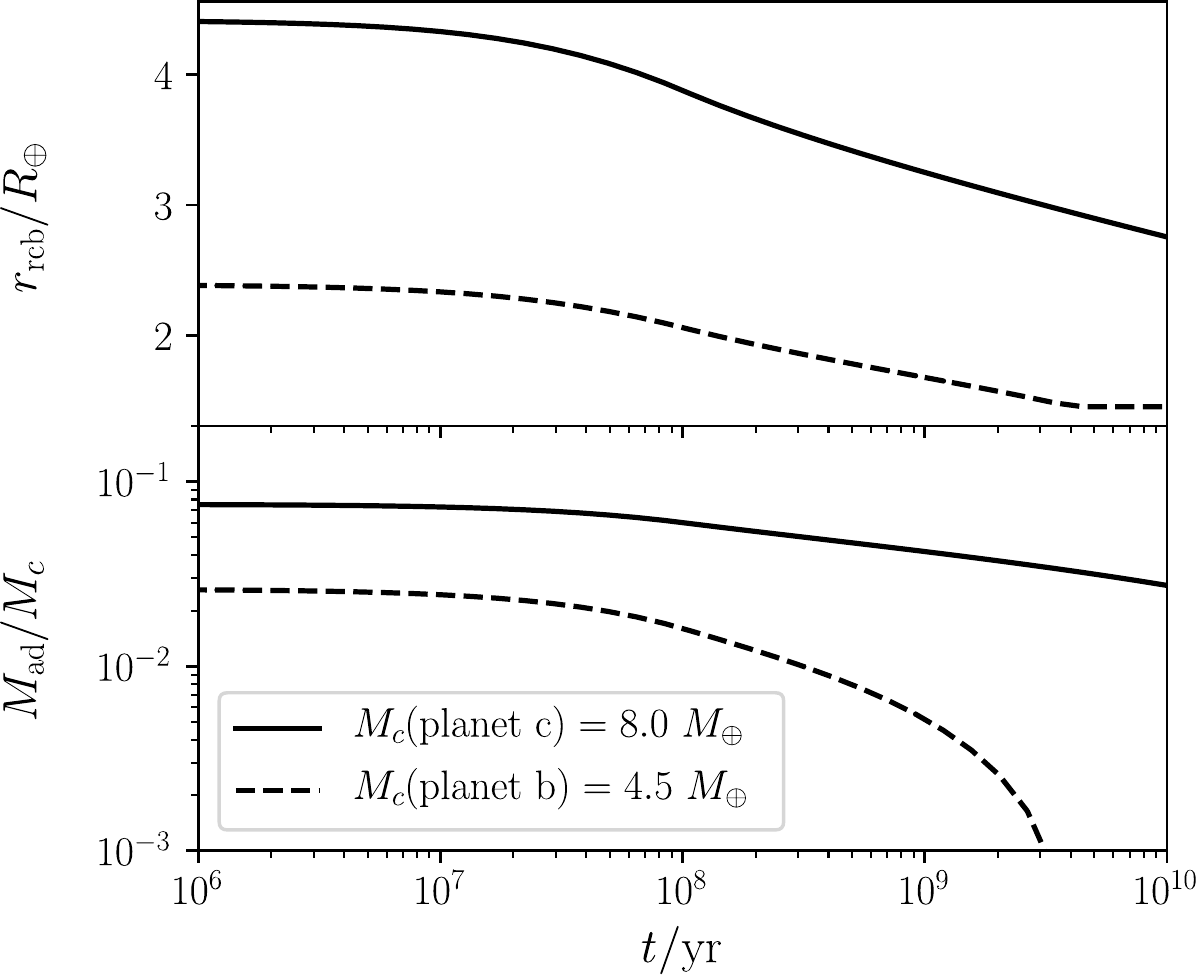}
  \caption{Evolution of model planet atmospheres, resembling
    the conditions of the two planets in Kepler-36, showing
    $r_\rcb$ and envelope mass fraction.  Note that the two
    planets start from the same initial envelope mass
    fraction, but the less massive one (corresponding to
    Kepler-36b) has already evaporated most of its envelope
    via Parker wind before $t = 10^6~\yr$.  See
    \S\ref{sec:sample-planet-evo} for detailed discussions.}
  \label{fig:kepler36}
\end{figure}

Equipped with eqs. \eqref{eq:mdot-fitting} and
\eqref{eq:reuv-estimate}, we can model the evolution tracks
of planet models by setting up initial conditions and then
integrate the ODE $\d M_\ad/\d t = \dot{M}$ for each model
planet.

Figure \ref{fig:t_diss_example} implies that, even the
external conditions and the initial conditions are nearly
the same, two planets would experience substantially
different tracks of evolution if they have different core
masses. Such systems are suggested by observations, but the
most clear detection is Kepler-36, whose two planets
circulate it at similar orbital radius ($\sim 0.12~\au$) but
have different core masses and observed radii (Kepler-36b:
$4.5~M_\oplus$, $1.49~R_\oplus$; Kepler-36c: $8~M_\oplus$,
$3.7~R_\oplus$; see also
\citealt{2012Sci...337..556C}). Here we setup two model
planets according to Kepler-36. For other parameters, we set
Kelvin-Helmholtz timescale $\tau_\kh = 10^8~\yr$ ,
isothermal layer temperature $T_\eq=850~\K$, and a evolving
high energy luminosity with initial value one order of
magnitude lower than the fiducial,
$L_\he(t) = 10^{-4.5}~L_\odot\times
\min\{1,(t/10^8~\yr)^{-1.5}\}$ (for the time-depence see the
discussions in \S\ref{sec:planet-ensembles}). Both planets
have initial envelope mass fraction
$M_\ad/M_c = 2\times 10^{-1}$ (as both models will lose
their atmosphere rapidly until they reach the pressure
balance line, this parameter is not sensitively depended
upon), and are integrated along the evolution tracks to
$10^{10}~\yr$. Figure \ref{fig:kepler36} show their fate: at
the age of Kepler-36, ($\sim 7~\mathrm{Gyr}$), the more
massive planet still has a relatively rich atmosphere
($M_\ad/M_c$ is few per cent), while the other has already
evaporated all its atmosphere. Considering the
$\sim (+0.5~R_\oplus)$ correction applied to the more
massive planet for the observed radius (given its $r_\rcb$
and core mass, e.g. \citealt{2014ApJ...792....1L}), the
bigger one should have $\sim 3.6~R_\oplus$ observed radius,
while the other is a $\sim 1.5~R_\oplus$ ``bare'' planet. We
thus conclude that the diverged evolution tracks due to
photoevaporation agrees with observation semi-quantitatively
(similar conclusion has also been reached by
e.g. \citealt{2013ApJ...776....2L, 2016ApJ...819L..10O}).

\subsubsection{Planet ensembles}
\label{sec:planet-ensembles}

\begin{figure}
  \centering
  \includegraphics[width=3.3in, keepaspectratio]
  {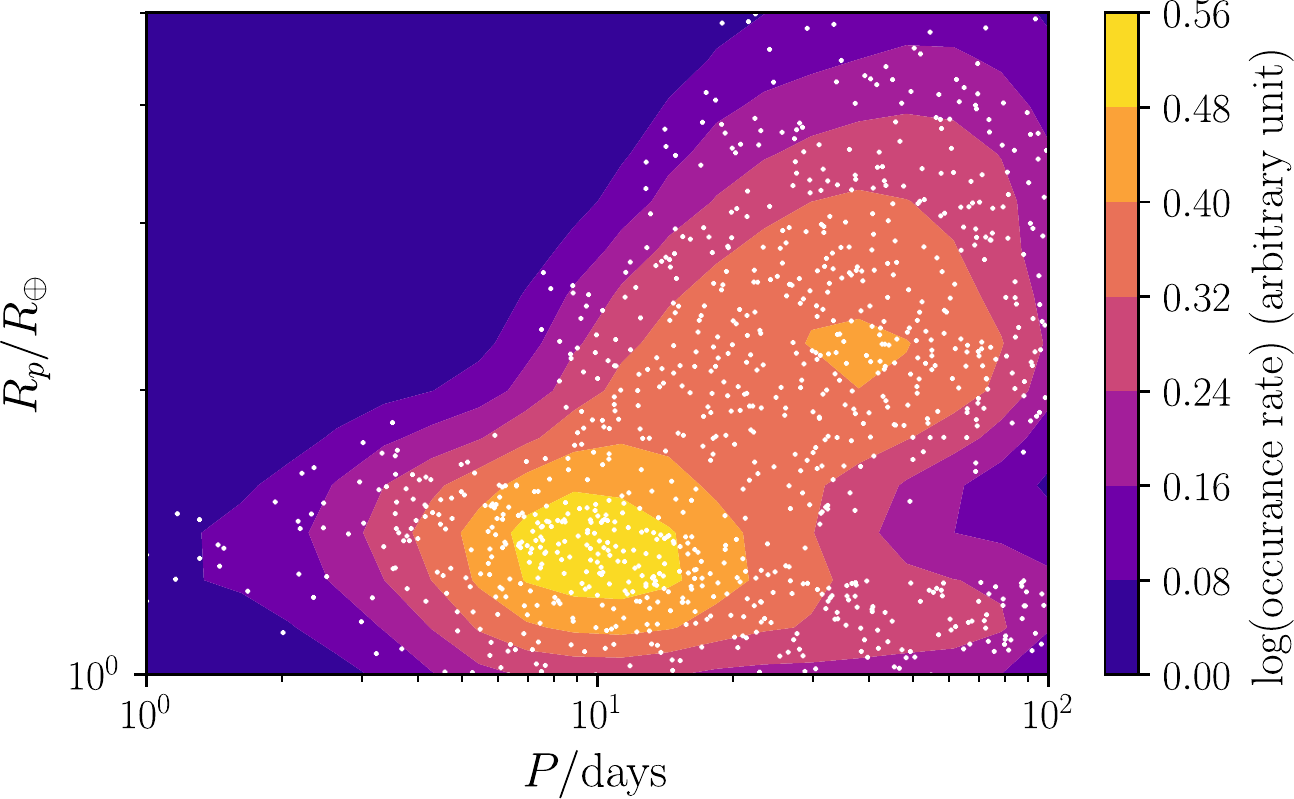}
  \caption{Bimodal distribution of evolved model
    planets. Scattered white dots present a sub-sample of
    simulated planet atmosphere evolution results ($10^3$
    out of $10^4$; for clearer presentation). The
    color-mapped contour is the estimated distribution
    function with Gaussian kernel, scaled to frequency per
    $\{\log_{10}{P}\times \log_{10}{R_p}\}$, using the
    results of the whole ensemble. }
  \label{fig:bimodal_dist}
\end{figure}

This subsection extends the calculation in
\S\ref{sec:sample-planet-evo} to a relatively big ensemble
of planets. We start by setting up the initial and input
conditions for the ensemble.

The distribution in orbital period is well-constrained by
Kepler observations. We adopt the recipes of OW17, based on
the observation results in \citet{2013ApJ...766...81F},
$\d N/\d \ln P\proptosim \min\{(P/7.6~\mathrm{days})^{1.9},\
1\}$, in the range $(P/\mathrm{days})\in[1,10^2]$. The
orbital periods are translated into orbital radii assuming
the host stellar mass being $M_\odot$. At each orbital
radius, $T_\eq$ and $F_\he$ are set accordingly. The core
mass distribution function is a Rayleigh function,
$\d N/\d M_c\proptosim M_c~\exp[-M_c^2/(2\sigma_M^2)]$ with
$\sigma_M\sim 3~M_c$ and $(M_c/M_\oplus)\in [1, 12]$
\citep[see also][]{2014ApJS..210...20M}. We assume that the
core of all model planets are rocky. The initial
distribution of envelope mass fraction, however, is
impossible to directly determine. We hence assume that the
distribution function of $\ln(M_\ad/M_c)$ is uniform in
$(M_\ad/M_c)\in [10^{-2},10^{-0.5}]$, and zero
elsewhere. All those distributions are assumed to be
independent to each other.

When evolving the planet envelopes, we integrate the mass
loss process to $10^9~\yr$. \citet{2005ApJ...622..680R}
suggested that the high energy luminosity of a young star
drops as power-law, whose power indices vary from band to
band. Here for simplicity, we assume that the power index is
the same $L\propto \min\{1,\ (t/10^8~\yr)^{-1.5}\}$ in all
bands, before totally shutting down radiation at $10^9~\yr$.
For those planet envelopes in the Parker wind zone in Figure
\ref{fig:t_diss_example}, we assume that it drops vertically
down to the photoevaporative region immediately before
evolving them photoevaporatively if $M_c> 3~M_\oplus$ , or
that it loses all envelope otherwise.

One possible caveat is the evolution of specific entropy in
the adiabatic interior of planet atmosphere, characterized
by the Kelvin-Helmholtz timescale, $\tau_\kh$. OW17
suggested that $\tau_\kh=\max\{10^8~\yr,\ t\}$ at time $t$
during the evolution process. Intuitive as it seems, we
realize that this recipes overestimates the overall energy
loss of planet envelopes after $10^8~\yr$, by comparing the
evolution of planet radii under OW17 scheme to the detailed
simulations, e.g. \citet{2015ApJ...808..150H}. Hence we
simply assume $\tau\simeq 10^8~\yr$ for all planets at all
times. This may overestimate observed planet radii at the
same $M_\ad$ after evaporation, but the post-evaporation
$M_\ad$ may be underestimated by having a bigger $R_\euv$
during the evolution, partially offsetting the former
overestimation.

By evolving the ensemble with $10^4$ model planets, we
obtain the frequency of evolved planet models on the
$\{\log_{10}{P}\times \log_{10}{R_p}\}$ plane (here $R_p$ is
characterized by $r_\rcb$), presented in Figure
\ref{fig:bimodal_dist}. The distribution is bimodal: one
peak locates at $R_p\sim 1.4~R_\oplus$,
$P\sim 10~\mathrm{days}$, and another at
$R_p\sim 2.5~R_\oplus$ , $P\sim 30~\mathrm{days}$.
Locations of these two peaks agree semi-quantitatively with
observation results in e.g. \citet{2017AJ....154..109F,
  2017arXiv170607807D}. Similar to OW17, this bimodality is
attributed to the peak in evolution timescales at
$M_\ad/M_c\sim 10^{-2}$, which is hereby confirmed in
detailed simulations with hydrodynamics, radiative transfer
and thermochemistry involved. The observations also confirm
the lack of objects at the upper left corner in the plot,
which is a direct result of photoevaporation
\citep[e.g.][]{2016NatCo...711201L}.

\goodbreak
\section{Summary}
\label{sec:summary}

In conclusion, this work studies the photoevaporation
processes of planet atmosphere by combining 2.5-dimensional
axisymmetric full hydrodynamic simulations with consistent
thermochemistry and ray-tracing radiative transfer.

As the initial conditions of photoevaporation, static planet
envelopes require the atmospheric pressure at large radii
being balanced by the ambient; otherwise, the envelopes may
lose mass rapidly through Parker wind. We find that for
planet core mass $M_c\lesssim 3~M_\oplus$, such balancing is
almost impossible to achieve by ambient pressure, which
suggests that they may not hold substantial H/He envelopes.
Semi-analytic models with spherical symmetry and hydrogen
ionization/recombination only suffer from the lack of
microphysics and proper hydrodynamics.

Numerical simulations reveal that the wind escape at
$32~\km~\s^{-1}$ with the ``standard'' high energy radiation
prescribed by \citet{2005ApJ...622..680R} and OW17, for a
planet with a $5~M_\oplus$ rocky core and envelope mass
fraction $10^{-2}$. Such a model planet loses its envelope
mass at $\dot{M}\simeq 4\times
10^{-10}~M_\oplus~\yr^{-1}$. While the outflow is fairly
close to radial on the day hemisphere, there exists a static
tail on the other side, shaped hydrodynamically by the flows
near $\theta = \pi/2$. We emphasize the importance of
multidimensionality by models whose supersonic outflows
still survive under strong stellar windram pressure, with
comparable mass loss rate to the fiducial case. By turning
off radiation flux in different bands, we find that the main
determinant of mass loss rate is the EUV photons, which
interact with most abundant species, H/\chem{H_2}/He. Other
bands of radiation assists the EUV photons by enlarging the
effective interception area of EUV and destroying molecular
and atomic coolants. Varying the planet and atmosphere
properties confirm that the size of EUV photosphere is the
most relevant factor. By setting different incident high
energy fluxes, we find that the mass loss rate drops
sub-linearly and the kinetic energy super-linearly at low
irradiation flux, where the mechanism limiting the outflow
is molecular ro-vibrational cooling rather than
recombination. Our numerical grid of planet models suggests
that the decrease of occurrence in rich planetary
atmospheres with core mass $M_c\lesssim 6~M_\oplus$ is
likely to be attributed to photoevaporation. Hinted by
numerical explorations, we propose a semi-empirical analytic
formula for the mass loss rate of photoevaporation, being
accurate to $\sim 20~\%$ in most cases, enabling further
viable predictions, e.g. simulating the evolution of planet
ensembles. We semi-quantitatively reproduce the bimodal
distribution of Kepler planet on the
$\{\log_{10}{P}\times \log_{10}{R_p}\}$ plane by evolving
such an ensemble, owing to the longest $t_\evap$ for
envelope mass fraction around $10^{-2}$.

In future works, we hope to explore the problem using models
with consistent thermochemistry in three dimensions. After
leaving the planet, photoevaporative outflow is subject to
modulations by orbital motion. Orbital centrifugal force and
the Coriolis force, which are expected to play a role even
within the Hill sphere of a planet in the co-rotational
frame, break the axisymmetry assumed in the 2.5-dimensional
models in this paper. Specifically, three dimensional models
will help us understand the behavior of the tail, which is
particularly interesting as similar structures are already
found in observations \citep[e.g.][]{2015Natur.522..459E,
  2017A&A...605L...7L}. Moreover, three dimensional
simulations will enable us to explore the interaction of
photoevaporative outflow with planet spin and even magnetic
field, leading to insights on more interesting physics
therein.

\vspace*{20pt} This work was partially supported by
Princeton University's Department of Astrophysical Sciences.
We thank our colleagues Xuening Bai and Jeremy Goodman,
for discussions and for detailed comments on a
preliminary draft. 

\appendix

\section{Equations for semi-analytic models}
\label{sec:app-semi-ana}

\subsection{Hydrodynamic and microphysical equations}
\label{sec:ana-hydro-mic-eq}

Under spherical symmetry, the hydrodynamic equations read,
in steady state (thus $\partial_t\equiv 0$),
\begin{equation}
  \label{eq:ana-dyn-eqn}
  \begin{split}
    & \dfrac{1}{r^2}\partial_r(r^2 \rho u )= 0\ ;\quad
    \dfrac{1}{r^2} \partial_r(r^2 \rho u^2 ) = -\partial_r 
    p - \dfrac{G M_c\rho}{r^2}\ ;
    \\
    & \dfrac{1}{r^2} \partial_r \left[ r^2 u \left(\gamma E
        + \dfrac{\rho u^2}{2} - \dfrac{G M_c\rho}{r} \right)
    \right] = S_E\ ,
  \end{split}
\end{equation}
where $u$ is the radial velocity, $E = p/(\gamma - 1)$ the
internal energy density of gas. Note that the gas pressure
satisfies $p=(1+x_e)\kb T n$, $n \equiv \rho/m_p$ is the
number density of hydrogen nuclei, $x_e \equiv n_e/n$ is the
ionized fraction. $S_E$ is the energy source term,
\begin{equation}
  \label{eq:ene-src}
  S_E = F \sigma n (1 - x_e) (h \nu - I_e) -
  \alpha_\B n^2 x_e^2 \mean{E_\rr} \ ,
\end{equation}
in which the radiation flux $F$ obeys the Lambert-Beer's
law,
\begin{equation}
  \label{eq:rad-flux}
  \partial_r \ln F = \sigma n (1 - x_e)\ .
\end{equation}
The conservation of elements, combined with mass
conservation, yields,
\begin{equation}
  \label{eq:ionization}
  u\partial_r x_e = S_I \equiv
  F \sigma (1 - x_e) - \alpha_\B n x_e^2 \  .
\end{equation}
By specifying reference dimensional variables $l_0$,
$\rho_0$ $c_{s 0}$, and $F_0$, the following equalities
define the dimensionless parameters (note that $\Theta$ and
$\epsilon$ are {\it not} independent; the Greek letters of
dimensionless quantities here are not to be confused with
physical quantities in other sections):
\begin{equation}
  \begin{split}
    & \lambda \equiv \dfrac{r}{l_0} \ , \
    \varrho \equiv  \dfrac{\rho}{\rho_0} \ , \
    \Theta \equiv \dfrac{T}{T_0} \ , \
    \mu \equiv \dfrac{u}{c_{s 0}} \ ,\
    \Gamma \equiv \dfrac{G M_c}{l_0^2c_{s0}^2}\ ,
    \delta \equiv \dfrac{\Delta}{c_{s 0}^2} =\mu^2 -
    \epsilon \varrho^{-1} \ , \
    \epsilon \equiv \dfrac{\gamma(\gamma -1)E}
    {c_{s0}^2\rho_0} \ ,\ 
    \varphi \equiv \dfrac{F}{F_0} \ ,
    \\
    & \xi_0 \equiv \dfrac{\mean{E_\rr}_{T=T_0}}{\kb T_0} = 
    \dfrac{3}{2} \ , \
    \zeta_0 \equiv \dfrac{h \nu - I_e}{I_e} \ , \
    \tau_0 \equiv \dfrac{\sigma \rho_0 l_0}{m_p}\ ,
    \\
    & C_1 \equiv \dfrac{F_0 \sigma I_e l_0}{c_{s 0}^3 m_p} \ ,\ 
    C_2 \equiv \alpha_0 \left( \dfrac{\rho_0}{m_p} \right)^2
    \left(\dfrac{\kb T_0 l_0}{\rho_0 c_{s 0}^3}
    \right) \ ,\ 
    C_3 \equiv \dfrac{F_0 \sigma l_0}{c_{s 0}} \ ,\
    C_4 \equiv \dfrac{\alpha_0 l_0 \rho_0}{c_{s 0} m_p} \ .
  \end{split}
\end{equation}
Then eqs. \eqref{eq:ana-dyn-eqn} through
\eqref{eq:ionization} are recast in their dimensionless
form,
\begin{equation}
  \label{eq:dimless-chem-dyn}
  \begin{split}
    \partial_\lambda \varrho
    & = (\mu \delta)^{- 1} \left[ (\gamma - 1) \Sigma_E
      - 2 \mu^3 \varrho / \lambda
      + \Gamma \mu \varrho / \lambda^2 \right] \  ;
    \\
    \partial_\lambda \mu
    & = -(\varrho \delta)^{-1} \left[ (\gamma - 1) \Sigma_E
      -2 \mu \epsilon / \lambda +
      \Gamma \mu \varrho / \lambda^2
      \right]\  ;
    \\
    \partial_\lambda \epsilon & =
    \delta^{- 1} \gamma \left[  (\gamma-1) \mu \Sigma_E
      - 2 \mu^2\epsilon / \lambda +  \Gamma
      \epsilon / \lambda^2 \right]\  ;
    \\
    \partial_\lambda x_e & = \mu^{-1} \Sigma_I \  ;
    \\
    \partial_\lambda \varphi & = - \tau_0 \varrho (1 - x_e)
    \varphi \ ;
    \\
    \Sigma_E & \equiv C_1 \varphi \varrho (1 - x_e) \zeta_0
    - C_2 \Theta^{\kappa + 1} x_e^2 \varrho^2 \xi_0 \  ;
    \\
    \Sigma_I & \equiv C_3 \varphi (1 - x_e) - C_4 
    \Theta^{\kappa} x_e^2 \varrho \  .
  \end{split}
\end{equation}

\subsection{Critical and boundary conditions}
\label{sec:ana-crit-bound-cond}

The ordinary differential equations (ODEs) in
eqs. \eqref{eq:dimless-chem-dyn} are singular at
$\delta = 0$, i.e. the radial velocity becomes
transonic. Physically feasible solutions (eigen solutions)
must pass through the sonic surface regularly by also having
vanishing numerators for $\partial_\lambda \varrho$,
$\partial_\lambda \mu$ and $\partial_\lambda \epsilon$. It
is straightforward to prove that numerators of those
derivatives vanish simultaneously if one of them approaches
zero as $\delta \rightarrow 0$. Near the sonic surface the
approximated derivatives are obtained by the l'Hospital
rule. Also, $\mu\rightarrow 0$ leads to singularity where
$\Sigma_E$ is still finite. We construct the solution by
starting at a finite radius
$\lambda_\mathrm{ini} = (r_\mathrm{ini} / l_0)$, then
integrate both inwards to $\lambda_\min = (r_\min / l_0)$,
defined as the wind base, where the dimensionless radiation
flux $\varphi =10^{-4}$, and outwards to
$\lambda_\max = (r_\max / l_0) = 10
\lambda_\mathrm{ini}$. The dependent variables at
$r_\mathrm{ini}$ are adjusted, so that the solution (a) is
regular at the sonic surface; (b) satisfies $\varphi = 1$ at
$r_\max$ by setting $F_0$ the incident EUV flux; and (c)
matches $\rho$ and $T$ of the given isothermal hydrostatic
profile at $r_\min$ (see also eq. \ref{eq:rho_isothermal}).

We are not matching $x_e$ at the inner boundary
$\lambda_\min$, as we do not prescribe the ionization
profile in the static region, while $x_e$ drops rapidly to
zero near and below $\lambda_\min$.  Comparing the number of
effective constraints ($\rho$ and $T$ at $r_\min$,
$\varphi = 1$ at $r_\max$, and regularity at the sonic
point; 4 constraints in total) to the number of dependent
variables ($\varrho$, $\mu$, $\epsilon$, $x_e$, and
$\varphi$; 5 dependent varialbes in total), the ODEs are
actually {\it underdetermined} with one degree of freedom:
with identical internal isothermal profile and external
radiation, a series of solutions (that are regular at
transonic points) can match the isothermal profile at
different $r_\min$. This mathematical consideration actually
has its physical implication, as is elaborated in
\S\ref{sec:static-caveat}.

\bibliography{planet_evap}
\bibliographystyle{apj}

%
\end{document}